\documentclass[11pt,oneside,reqno]{amsart}
\usepackage[T1]{fontenc}
\usepackage[latin9]{inputenc}
\usepackage[a4paper]{geometry}
\usepackage{float}
\usepackage{amstext}
\usepackage{amsthm}
\usepackage{amssymb}
\usepackage{setspace}
\usepackage[authoryear]{natbib}
\usepackage{color}
 \usepackage{bbm}
\makeatletter

\providecommand{\tabularnewline}{\\}

\numberwithin{equation}{section}
\numberwithin{figure}{section}

\usepackage{amsfonts}
\usepackage{amstext}
\usepackage{amsthm}

\newtheorem{thm}{Theorem}
\newtheorem{lem}{Lemma}

\newtheorem{asm}{Assumption}

\theoremstyle{definition}
\newtheorem{rem}{Remark}

\newcommand{\suml}{\sum_{l=1}^n }

\makeatother

\begin{document}
\title{Multiway empirical likelihood}
\author{Harold D. Chiang}
\address{Department of Economics, University of Wisconsin-Madison, 1180 Observatory Drive Madison, WI 53706-1393, USA.}
\email{hdchiang@wisc.edu}
\author{Yukitoshi Matsushita}
\address{Graduate School of Economics, Hitotsubashi University, 2-1 Naka, Kunitachi, Tokyo 186-8601, Japan.}
\email{matsushita.y@r.hit-u.ac.jp}
\author{Taisuke Otsu}
\address{Department of Economics, London School of Economics, Houghton Street, London, WC2A 2AE, UK.}
\email{t.otsu@lse.ac.uk}
\thanks{We extend our gratitude to the associate editor and the three anonymous referees for their valuable and constructive feedback, which has greatly enhanced the quality of our paper. We thank the participants at Cowles Econometrics Conference 2022, NY Camp Econometrics XVI, SEA2022, Workshop on Specification and Misspecification Analysis in Cross-Section and Panel Data at UvA, and seminars participants at Brown, Columbia, Indiana-Bloomington, Princeton, NCSU, Northwestern, Singapore Management U, TAMU, UC Irvine, UCLA, UChicago, and UWashington-Seattle for their valuable comments and discussions.  H. Chiang is supported by Office of the Vice Chancellor for Research and Graduate Education at UW-Madison with funding from Wisconsin Alumni Research Foundation. Y. Matsushita is partly supported by the JSPS KAKENHI grant number 18K01541.}
\begin{abstract}
This paper develops a general methodology to conduct statistical inference for observations indexed by multiple sets of entities. We propose a novel multiway empirical likelihood statistic that converges to a chi-square distribution under the non-degenerate case, where corresponding Hoeffding type decomposition is dominated by linear terms. Our methodology is related to the notion of jackknife empirical likelihood but the leave-out pseudo values are constructed by leaving out columns or rows. We further develop a modified version of our multiway empirical likelihood statistic, which converges to a chi-square distribution regardless of the degeneracy, and discuss its desirable higher-order property in a simplified setup. The proposed methodology is illustrated by several important econometric problems, such as bipartite network, generalized estimating equations, and three-way observations. 
\end{abstract}

\maketitle

\allowdisplaybreaks
\section{Introduction\label{sec:intro}}

Many important econometric problems feature multiway data, in which observations are indexed by multiple sets of entities, often arranged as rows and columns. Examples include longitudinal data \citep{liang1986longitudinal}, classical random effect models \citep[Ch 5]{searle2009variance}, row-column exchangeable models \citep{Mccullagh2000}, nonnested multilevel data \citep{miglioretti2007marginal}, bipartite networks \citep{choi2014co}, and multiway clustering \citep{CGM2011}, to list a few. 
Observations in such datasets, when correspond to a same set of entity, can exhibit strong dependence that does not diminish as certain distance measure increases, which invalidates conventional asymptotic theory. Although Eicker-White type multiway cluster robust standard errors have been developed for statistical inference on multiway data, and are frequently used in empirical research,\footnote{For example, according to Google Scholar, as of 13th of October, 2021, \citet{CGM2011} receives over $3,100$ citations, while \citet{thompson2011simple} has over $1,400$ citations.} (i) their derivations are largely case-by-case, (ii) the resulting inference may not be reliable in finite samples, especially when one of the index dimensions contains only a moderate number of units,  (iii) in situations with weak or no cluster dependence, the resulting inference often demonstrates significantly less precision than in the cases with strong dependence, and
(iv) they often underestimate the variance in finite samples and lead to distortions in the size or coverage properties.

This paper develops a general framework to conduct inference on statistical models for various multiway data. In particular, inspired by the idea of jackknife empirical likelihood of \cite{JingYuanZhang2009}, we propose a novel multiway empirical likelihood (MEL). Unlike the conventional leave-one-out operation for jackknifing, one leaves all the observations in a column or a row out at a time to construct the leave-out pseudo values. The resulting MEL function is computationally attractive and is shown to be asymptotically pivotal when the linear terms of the Hoeffding type decomposition of the statistical object of interest dominate the quadratic terms (called the non-degenerate case). In multiway data and models, however, degeneracy often occurs so that the quadratic terms in the Hoeffding type decomposition emerge in the first-order, and the MEL statistic loses its asymptotic pivotalness. This phenomenon can be understood as an analogy of emergence of the Efron-Stein bias for the jackknife variance estimator in the multiway context even though in the original setup of \cite{EfronStein1981}, the bias is of second-order. To recover asymptotic pivotalness, we modify the baseline MEL statistic by incorporating leave one column and row out adjustments, which may be considered as an extension of the leave-two-out bias correction idea in \cite{hinkley1978improving} and \cite{EfronStein1981} to our two-way setup. Under mild regularity conditions, this modified MEL statistic converges to a chi-square distribution regardless of the degeneracy.

To further motivate our modified MEL approach, we investigate higher-order properties of the modified MEL statistic in a simplified setup. In this analysis, we find that the second-order term of the asymptotic expansion for the modified MEL statistic is closer to zero than that of the Wald statistic with the Eicker-White type robust standard error, which is always negative. Although the results are derived under a restrictive setting, our higher-order analysis illustrates an advantage of our modified MEL inference and also provides some explanation on the oversize phenomenon of the Wald test based on Eicker-White cluster robust standard errors that has been well-documented in the literature \citep[for example]{CGM2011,thompson2011simple,mackinnon2021wild}. 

We illustrate wide-applicability of the modified MEL method by various econometric applications, including sparse bipartite network formation models \citep{BickelChenLevina2011, Graham2020logit}, and generalized estimating equations \citep{liang1986longitudinal, xie2003asymptotics, balan2005asymptotic}. Throughout these different contexts, the proposed methodology can be applied without modification. We also generalize the modified MEL method to a three-way index setting. Finally, we conduct simulation studies over various settings for classical random effect models and bipartite stochastic block models. The results suggest that the finite sample performance of the modified MEL significantly dominates the Eicker-White multiway cluster robust standard error. The difference is especially profound when cluster dependence is weak or absent. Furthermore, in contrast to the Eicker-White procedure, the modified MEL delivers reliable coverage probabilities even when one of the index dimensions contains only a moderate number of observations.

This paper also contributes to the literature of empirical likelihood (\cite{Owen1988}; see \cite{Owen2001Book} for an overview). After the seminal work by \cite{JingYuanZhang2009}, jackknife empirical likelihood and its variants have been extended to various econometric and statistical problems, e.g., \cite{GongPengQi2010JMA}, \cite{zhang2013empirical}, \cite{zhong2014jackknife}, among others. In particular, \cite{MatsushitaOtsu2020} proposed modified jackknife empirical likelihood to cope with the Efron-Stein bias and established its asymptotic pivotalness under both conventional and non-standard asymptotics. Under the conventional asymptotics, empirical likelihood inference has been studied and extended to various contexts; see e.g., \cite{bertail2006empirical}, \cite{zhu2006empirical}, \cite{HjortMcKeagueVan_Keilegom2009AoS}, \cite{BravoEscancianoVan_Keilegom2020AoS}, and a review by \cite{ChenVan_Keilegom2009Test}, among many others.

This paper is organized as follows. Section \ref{sec:bench} presents our basic theoretical results on the MEL and its modification for inference on the means of two-way data. Sections \ref{sub:1st} and \ref{sub:2nd} study the first and higher-order asymptotic properties, respectively. In Section \ref{sec:general}, we extend our MEL approach to a bipartite network model (Section \ref{sub:network}), generalized estimating equations (Section \ref{sub:GEE}), and three-way data (Section \ref{sub:three-way}). Section \ref{sec:sim} illustrates the proposed method by two simulation examples. All proofs are contained in the Appendix.

\section{Benchmark case: Two-way empirical likelihood for mean\label{sec:bench}}

\subsection{First-order asymptotic theory\label{sub:1st}}
As a benchmark, for each $N,M\in \mathbbm N$, we first consider a two-way sample of $d$-dimensional random vectors $\{X_{ij}:i=1,\ldots,N,j=1,\ldots,M\}$ generated by
\begin{equation}
X_{ij}=\tau(U_{i0},U_{0j},U_{ij}),\label{eq:AH}
\end{equation}
for $i=1,\ldots,N$ and $j=1,\ldots,M$, where $\{U_{i0},U_{0j},U_{ij}:i=1,\ldots,N,j=1,\ldots,M\}$ are i.i.d. unobservable latent shocks that can be normalized to $U[0,1]$,\footnote{Independence over $(U_{0j})_{j\in \mathbb{N}}$ is assumed for simplicity, and can be replaced by a martingale difference sequence type condition.} and $\tau=\tau_{(N,M)}$ is an unknown $\mathbb R^d$-valued Borel-measurable map that may vary with $(N,M)$. Hereafter all population objects, such as the distribution and moments of $X_{ij}$, depend on $(N,M)$ through $\tau(\cdot)$, but we suppress the dependence on $(N,M)$ for notational brevity. In this benchmark setup, we consider statistical inference on the mean vector $\theta=E[X_{11}]$ by using the MEL method.

A common set of sufficient conditions for the representation in (\ref{eq:AH}) is that $\{X_{ij}:i=1,\ldots,N,j=1,\ldots,M\}$  is dissociated and is embedded into an infinite two-way separately exchangeable array $(X_{ij})_{(i,j)\in\mathbb{N}^{2}}$. A set of random variables $\{X_{ij}: =1,...,N, j=1,...M\}$ is dissociated if for any two disjoint sets $A,B\subset \{1,...,N\}\times \{1,...,M\}$, $\{X_{ij}:(i,j)\in A\}$ and $\{X_{ij}:(i,j)\in B\}$ are independent. An infinite array $(X_{ij})_{(i,j)\in\mathbb{N}^{2}}$ is called separately exchangeable if for any two permutations of positive integers $\pi_{1},\pi_{2}:\mathbb{N}\to\mathbb{N}$ and any finite subset $A\subset\mathbb{N}^{2}$, it holds $(X_{ij})_{(i,j)\in A}\overset{d}{=}(X_{\pi_{1}(i)\pi_{2}(j)})_{(i,j)\in A}$. Under such conditions, (\ref{eq:AH}) follows from the celebrated Aldous-Hoover-Kallenberg representation for separately exchangeable arrays (e.g., Corollary 7.23 of \citealp{Kallenberg2006}). This representation is widely used in modern statistics, e.g., \citet{diaconis2008graph}, \citet{BickelChenLevina2011}, \citet{choi2014co}, \citet{bhattacharyya2015subsampling}, \citet{gao2015rate}, \citet{caron2017sparse}, \citet{choi2017co}, \citet{zhang2017estimating}, \cite{lauritzen2018random}, \citet{veitch2019sampling},  \citet{davezies2021}, \cite{mackinnon2021wild}, \citet{menzel2021bootstrap}, and many more. See also the review by \cite{orbanz2014bayesian}.

The representation in (\ref{eq:AH}) is useful for our theoretical development since it allows us to establish a Hoeffding type decomposition for the estimation error of the point estimator $\hat{\theta}=\frac{1}{NM}\sum_{i=1}^{N}\sum_{j=1}^{M}X_{ij}$, that is 
\begin{equation}
\hat{\theta}-\theta=\frac{1}{N}\sum_{i=1}^{N}L_{i0}+\frac{1}{M}\sum_{j=1}^{M}L_{0j}+\frac{1}{NM}\sum_{i=1}^{N}\sum_{j=1}^{M}(W_{ij}+R_{ij}),\label{eq:Hoeff}
\end{equation}
where 
\begin{eqnarray*}
L_{i0} & = & E[X_{i1}|U_{i0}]-E[X_{11}],\qquad L_{0j}=E[X_{1j}|U_{0j}]-E[X_{11}],\\
W_{ij} & = & E[X_{ij}|U_{i0},U_{0j}]-E[X_{i1}|U_{i0}]-E[X_{1j}|U_{0j}]+E[X_{11}],\\
R_{ij} & = & X_{ij}-E[X_{ij}|U_{i0},U_{0j}].
\end{eqnarray*}
Observe that $\{L_{i0}:i=1,\ldots,N\}$ and $\{L_{0j}:j=1,\ldots,M\}$ are i.i.d., and $\{R_{ij}:i=1,\ldots,N,j=1,\ldots,M\}$ is also i.i.d. conditional on $\{U_{i0},U_{0j}:i=1,\ldots,N,j=1,\ldots,M\}$.

Throughout this paper, we regard $S(\theta)=\hat{\theta}-\theta$ as an estimating equation for $\theta$ and construct the MEL function to conduct inference on $\theta$. More precisely, let us define $n=N+M$ and introduce the leave-out pseudo value:
\[
V_{l}(\theta)=nS(\theta)-(n-1)S_{l}(\theta),
\]
for $l=1,\ldots,n$, where $S_{l}(\theta)=\hat{\theta}^{(l)}-\theta$ and $\hat{\theta}^{(l)}$ is the leave one column or row out counterpart of $\hat{\theta}$ defined as 
{  \[
\hat{\theta}^{(l)}=\left\{ \begin{array}{cc}
\frac{1}{(N-1)M}\sum^{N}_{i=1, i \ne l}\sum_{j=1}^{M}X_{ij} & \text{if }l\le N,\\
\frac{1}{N(M-1)}\sum_{i=1}^{N}\sum^{M}_{j=1, j\ne l-N}X_{ij} & \text{otherwise.}
\end{array}\right.
\]}
Unlike the conventional leave-one-out operation for jackknifing, we leave all the observations that have a specific $i$ or $j$ out at a time. Thus the number of leave-out pseudo values is $n=N+M$ instead of $NM$, which indicates computational attractiveness of our MEL method.

By considering the leave-out pseudo value $V_{l}(\theta)$ as an estimating function for $\theta$, the MEL function for $\theta$ is constructed as 
\[
\ell(\theta)=-2\sup_{w_{1},\ldots,w_{n}}\sum_{l=1}^{n}\log(nw_{l})\qquad\text{s.t.}\quad w_{l}\ge0,\quad\sum_{l=1}^{n}w_{l}=1,\quad\sum_{l=1}^{n}w_{l}V_{l}(\theta)=0.
\]
{  Although this constrained optimization involves $n$ variables $\{w_{1}\ldots,w_{n}\}$, we can apply the Lagrange multiplier method to derive its dual form (see Section 3.14 of Owen, 2001), that is 
\begin{equation}
\ell(\theta)=2\sup_{\lambda\in \mathbb{R}^d}\sum_{l=1}^{n}\log(1+\lambda^{\prime}V_{l}(\theta)),\label{eq:MEL}
\end{equation}
where $\lambda$ is a vector of the Lagrange multipliers for the constraints $\sum_{l=1}^{n}w_{l}V_{l}(\theta)=0$, and  $\lambda'$ denotes the transpose of $\lambda$. In practice $\ell(\theta)$ can be computed by the dual form since the dimension of $\lambda$ is much lower than $n$.}

To study asymptotic properties of the MEL statistic $\ell(\theta)$, we impose the following assumptions. Let $\underline{n}=N\wedge M$, $\sigma_{L}^{2}=n\{Var(L_{10})/N+Var(L_{01})/M\}$ and {  $\sigma_{R}^{2}=Var(R_{11})$} be the variance matrices of the components in the Hoeffding type decomposition in (\ref{eq:Hoeff}), and $\lambda_{\min}(A)$ be the minimum eigenvalue of a matrix $A$. We say the sequence of data-generating processes is {\it non-degenerate} if $\lambda_{\min}(\underline{n}\sigma_{L}^{2})\to\infty$ and {\it nearly degenerate} if $\lambda_{\min}(\underline{n}\sigma_{L}^{2})=O(1)$.

\begin{asm} \label{asm:MEL}(i) $\{X_{ij}:i=1,\ldots,N,j=1,\ldots,M\}$ is generated by (\ref{eq:AH}).
(ii) For some $q>4$, $E[\|X_{11}\|^{q}]$ is bounded from above uniformly in $n$. Also $\lim_{n\to\infty}\underline{n}/(N\lor M)\in(0,1)$.
(iii) Under the nearly degenerate case, $\lambda_{\min}(\sigma_{R}^{2})\ge c>0$ for a constant $c$ independent of $n$ and $\|\sum_{i=1}^{N}\sum_{j=1}^{M}W_{ij}\|/\|\sum_{i=1}^{N}\sum_{j=1}^{M}R_{ij}\|=o_{p}(1)$.
\end{asm}

\begin{rem} Assumption \ref{asm:MEL}(i) assumes that the data have a two-way dependence structure. Assumption \ref{asm:MEL}(ii) requires the observables to have more than four moments, as well as limiting the growth rates of $N$ and $M$ to be similar. This can be loosen by imposing alternative assumptions to control the growth rates of $Var(L_{10})$ and $Var(L_{01})$. Assumption \ref{asm:MEL}(iii) imposes some high level conditions on the asymptotic behaviour of the term $W_{ij}$ in the Hoeffding type decomposition for the nearly degenerate case. It assumes that the term $R_{ij}$ in (\ref{eq:Hoeff}) remains random asymptotically, but the term $\sum_{i=1}^N \sum_{j=1}^M W_{ij}$ is asymptotically negligible compared to the $\sum_{i=1}^N \sum_{j=1}^MR_{ij}$ term. 
	{ 	
We emphasize that the latter condition in Assumption \ref{asm:MEL}(iii) is only binding in the near degenerate scenarios, and holds automatically true in the following four representative applications. First, suppose the latent components are additively separable (such as the classical random effect models in \citealt[Ch 5]{searle2009variance}), i.e.,
\begin{align*}
X_{ij}=a(U_{i0})+b(U_{0j}) + c(U_{ij}),
\end{align*}
for some $\mathbb R^d$-valued functions $a,b,c$. Then it holds $W_{ij}=0$ and the latter condition in Assumption \ref{asm:MEL}(iii) is satisfied. Second, suppose the latent shocks specific to $i$ and $j$ enter the observable random variable in an additively separable manner, e.g.,
\begin{align*}
X_{ij}=a(U_{i0},U_{ij})+b(U_{0j},U_{ij}),
\end{align*}
for some $\mathbb R^d$-valued functions $a,b$. Then again $W_{ij}=0$ and the latter condition in Assumption \ref{asm:MEL}(iii) is satisfied.
Third, consider asymptotics of sparse networks \citep{BickelChenLevina2011, Graham2020logit} such that the random variable of interest takes a binary value with a mean converging to zero:
\begin{align*}
X_{ij}= \mathbbm 1\{g_{MN}(U_{i0},U_{0j},U_{ij})>0\},\quad
E[X_{ij}]=\theta_{NM} =o(1),
\end{align*}
for a sequence of real-valued functions $g_{MN}$. Then we obtain
\begin{align*}
Var(W_{ij})=O(\theta^2_{NM})\ll Var(R_{ij})= O(\theta_{NM}),
\end{align*}
and the latter condition in Assumption \ref{asm:MEL}(iii) is satisfied. Finally, consider the kernel-type estimation of directed dyadic density models \citep{Graham2019kernel}, i.e., suppose $X_{ij}$ consists of the summand of a kernel density estimation problem
\begin{align*}
X_{ij}=\frac{1}{h} K\left(\frac{Y_{ij}-y}{h}\right),
\end{align*}
where the random variable of interest  $Y_{ij}=g(U_{i0},U_{0j},U_{ij})$ is an observed identically distributed random variable with the density $f_Y$ satisfying $f_Y(y)\ge c >0$, $K$ is a second order kernel function, $h=h_{NM}=o(1)$ is a sequence of bandwidth parameter satisfying $nh\to \infty$. Then under appropriate regularity conditions, one can show that
\begin{align*}
Var(W_{ij}) = O(n^{-1}) \ll Var(R_{ij})=O((nh)^{-1}),
\end{align*}
which guarantees the latter condition in Assumption \ref{asm:MEL}(iii).}
Finally, we remark that, at the cost of lengthier assumptions, one can relax this condition by applying Corollary 1 in \cite{chiang2023using} to allow $\sum_{i=1}^N \sum_{j=1}^M W_{ij}$  and $\sum_{i=1}^N \sum_{j=1}^MR_{ij}$ components to have the same stochastic order. \end{rem}

Under these assumptions, the limiting distribution of the MEL statistic $\ell(\theta)$ is obtained as follows.

\begin{thm} \label{thm:MEL} Under Assumption \ref{asm:MEL}, it holds 
\[
\ell(\theta)\overset{d}{\to}
\begin{cases}
\chi_{d}^{2} & \text{for non-degenerate case},\\
\xi^{\prime}\Omega^{-1}\xi & \text{for nearly degenerate case},
\end{cases}
\]
where $\xi\sim N(0,\lim_{n\to\infty}(\underline{n}\sigma_{L}^{2}+\sigma_{R}^{2}))$ and $\Omega=\lim_{n\to\infty}(\underline{n}\sigma_{L}^{2}+2\sigma_{R}^{2})$.
\end{thm}

\begin{rem}\label{rem:mMW} This theorem says that the asymptotic distribution of the MEL statistic $\ell(\theta)$ depends on the behaviour of the variance component $\sigma_{L}^{2}$ for the linear term in (\ref{eq:Hoeff}). If it is asymptotically non-negligible in the sense that $\sigma_{L}^{2}$ does not converge to zero at least as fast as $\underline{n}^{-1}$, then the MEL statistic is asymptotically pivotal. However, when $\sigma_{L}^{2}$ converges to zero at a rate of $\underline{n}^{-1}$ or faster, then the MEL statistic is no longer pivotal and its asymptotic distribution depends on $\underline{n}$, $\sigma_{L}^{2}$, and $\sigma_{R}^{2}$. If $\tau$, $(U_{i0})_i$, and $(U_{0j})_j$ in (\ref{eq:AH}) are known to the statistician and $(U_{ij})_{i,j}$ do not enter $\tau$, then the discrepancy between ``$2\sigma_{R}^{2}$'' in $\Omega$ and ``$\sigma_{R}^{2}$'' in the variance of $\xi$ can be understood as a two-sample $U$-statistic generalization of the second order Efron-Stein bias. Nonetheless, under our asymptotic framework, this bias emerges in the first-order. {  A similar phenomenon is also documented in Theorem 2 in \cite{mackinnon2021wild} for the ``two-term" version of Eicker-White type cluster robust variance estimator.} \end{rem} 

In order to conduct statistical inference based on the MEL statistic $\ell(\theta)$, we need to employ different critical values for the different cases. In particular, for the nearly degenerate case, we need to estimate $Var(\xi)$ and $\Omega$. Thus, it is desirable to modify the MEL statistic to have the same limiting distribution for both cases.

Motivated by the bias correction method in \citet{EfronStein1981}, we develop a modified version of the MEL statistic as follows. For each $l=1,\ldots,N$ and $l_{1}=1,\ldots,M$, let $\hat{\theta}^{(l,l_{1})}=\frac{1}{(N-1)(M-1)}\sum_{i\ne l}\sum_{j\ne l_{1}}X_{ij}$ be the leave one-column and one-row out counterparts of $\hat{\theta}$, $S_{l,l_{1}}(\theta)=\hat{\theta}^{(l,l_{1})}-\theta$, and 
\begin{equation}
Q_{ll_{1}}=\mathcal C(N,M)\cdot [nS(\theta)-(n-1)\{S_{l}(\theta)+S_{N+l_{1}}(\theta)\}+(n-2)S_{l,l_{1}}(\theta)], \label{eq:adjQ}
\end{equation}
where $\mathcal C(N,M)=\frac{(N-1)(M-1)n}{(NM)(n-2)}$. Note that the term $Q_{ll_{1}}$ is different from the leave-two-out counterpart employed in \citet{EfronStein1981} to correct the higher-order bias of the jackknife variance estimator since we delete a whole column and row of the data matrix $(X_{ij})$. Therefore, the total number of leave-out estimators required here is of order $O(n^2)$, similar to the usual leave-one-out procedures, in contrast to ${NM \choose 2}=O(n^4)$ of the conventional leave-two-out methods. The factor $\mathcal C(N,M)$ is a finite sample adjustment to make the coefficient of the leading term $(W_{ll_{1}}+R_{ll_{1}})$ of $Q_{ll_{1}}$ to be ``$\frac{n}{NM}$'' as in (\ref{pf:Qll1}) in the Appendix so that the leading term of the adjustment term $\frac{MN}{n^{2}}\sum_{l=1}^{N}\sum_{l_{1}=1}^{M}Q_{ll_{1}}Q_{ll_{1}}^{\prime}$ will be $\frac{1}{NM}\sum_{l=1}^{N}\sum_{l_{1}=1}^{M}(W_{ll_1}+R_{ll_1})^{2}$, which is unbiased for $Var(W_{ll_1}+R_{ll_1})$.

Based on $Q_{ll_{1}}$, the modified estimating function is defined as 
\[
V_{l}^{m}(\theta)=V_{l}(\hat{\theta})-\hat{\Gamma}\tilde{\Gamma}^{-1}\{V_{l}(\hat{\theta})-V_{l}(\theta)\},
\]
for $l=1,\ldots,n$, where $\hat{\Gamma}$ and $\tilde{\Gamma}$ are so that
\begin{equation}
\hat{\Gamma}\hat{\Gamma}^{\prime}=\frac{1}{n}\sum_{l=1}^{n}V_{l}(\hat{\theta})V_{l}(\hat{\theta})^{\prime},\qquad { \tilde{\Gamma}\tilde{\Gamma}^{\prime}=\frac{1}{n}\sum_{l=1}^{n}V_{l}(\hat\theta)V_{l}(\hat\theta)^{\prime}-\frac{1}{n}\sum_{l=1}^{N}\sum_{l_{1}=1}^{M}Q_{ll_{1}}Q_{ll_{1}}^{\prime}}.\label{eq:correct}
\end{equation}
By using $V_{l}^{m}(\theta)$ as a moment function, the modified MEL statistic is defined as 
\begin{equation}
\ell^{m}(\theta)=2\sup_{\lambda\in \mathbb{R}^d}\sum_{l=1}^{n}\log\{1+\lambda^{\prime}V_{l}^{m}(\theta)\},\label{eq:mMEL}
\end{equation}
and the asymptotic property of this statistic is obtained as follows.

\begin{thm} \label{thm:mMEL} Under Assumption \ref{asm:MEL}, it holds (for both non-degenerate and nearly degenerate cases) 
\[
\ell^{m}(\theta)\overset{d}{\to}\chi_{d}^{2}.
\]
\end{thm}

This theorem shows that the modified MEL statistic $\ell^{m}(\theta)$ has the asymptotically pivotal distribution of $\chi_{d}^{2}$ for both asymptotic regimes. We emphasize that the modified MEL inference only requires the estimators, $\hat{\theta}$, $\hat{\theta}^{(l)}$, and $\hat{\theta}^{(l,l_{1})}$, and circumvents estimation of $Var(\xi)$ and $\Omega$ in Theorem \ref{thm:MEL}. Based on this theorem, the asymptotic $100(1-\alpha)\%$ modified MEL confidence set can be constructed as $\{\theta:\ell^{m}(\theta)\le\chi_{d,\alpha}^{2}\}$, where $\chi_{d,\alpha}^{2}$ is the $(1-\alpha)$-th quantile of the $\chi_{d}^{2}$ distribution.

\begin{rem}\label{rem:mMW}
A by-product of the proposed modified MEL procedure is the modified multiway variance estimator $\tilde{\Gamma}\tilde{\Gamma}^{\prime}$ in (\ref{eq:correct}) evaluated at $\theta = \hat{\theta}$, an alternative to the Eicker-White multiway cluster robust variance estimators. It can be considered as an analogy of the bias-corrected jackknife variance estimator \citep{EfronStein1981} for our multiway context. Based this variance estimator, we can also construct a confidence interval $\left[\hat{\theta}_{j}\pm n^{-1/2} z_{\alpha/2}\sqrt{[\tilde{\Gamma}\tilde{\Gamma}^{\prime}]_{(j,j)}}\right]$ for the $j$-th element $\theta_{j}$ of $\theta$. This confidence interval seems to be new in the literature, and in contrast to \cite{EfronStein1981}, the correction term in the new standard error $\sqrt{[\tilde{\Gamma}\tilde{\Gamma}^{\prime}]_{(j,j)}}$ is not asymptotically negligible in the first-order. {  In addition, in the (completely) degenerate case of $\sigma_{L}=0$, the implied convergence rate of the mean of this modified estimating function $V_l^m(\theta)$ is $\sqrt{NM}\sim n$ rather than the slower rate of $\sqrt{n}$. Thus, the proposed modification does not slow down the convergence rate of distributional approximation of the test statistic. }
\end{rem}

\subsection{Higher-order properties\label{sub:2nd}}

In this subsection, we provide some theoretical justification for desirable accuracy of the modified MEL statistic by the asymptotic $\chi^{2}$ calibration based on the higher-order property {  in a simplified setup:
\begin{equation}
X_{ij}=\theta + \varepsilon_{ij}, \label{eq:simple}
\end{equation}
where $\{\varepsilon_{ij}\}$ is a scalar sequence of i.i.d. random variables\footnote{ As such, the results in Theorem 3 are not directly comparable with the classical Edgeworth expansion results for U-statistics, such as those in \cite{helmers1991edgeworth}, \cite{putter1998empirical}, etc.} for $i=1,\ldots,N$ and $j=1,\ldots,M$. In this particular scenario, our simulation studies below present substantial disparities in performance between the modified MEL method and the conventional inference procedure relying on the Eicker-White variance estimator. Although this setup covers some specific examples, such as the random effect model with $\sigma^2=0$ in the simulation study below and the Erd\H{o}s-R\'{e}nyi model, it is arguably restrictive. In the context of network data analysis, a recent paper by Zhang and Xia (2022) thoroughly studied higher-order properties of the network moments via novel Edgeworth expansion techniques. As clarified in Zhang and Xia (2022), the network moment statistics are considered as noisy U-statistics that involve edgewise errors to observe the adjacency matrix, and their higher-order analysis is substantially different from the conventional noiseless case. It is beyond the scope of this paper to extend Zhang and Xia's (2022) higher-order analysis to the modified MEL and Wald statistics under our multiway setting.}

We compare the second-order terms of the distributions of the modified MEL statistic $\ell^{m}(\theta)$ and the Wald statistic or $t$-ratio based on the Eicker-White type cluster robust variance estimator, i.e., $T(\theta)=(\hat{\theta}-\theta)^{\prime}\hat{\Sigma}^{-1}(\hat{\theta}-\theta)$, where 
\begin{eqnarray*}
\hat{\Sigma} & = & \frac{1}{N^{2}M^{2}}\sum_{i=1}^{N}\sum_{j=1}^{M}\sum_{j_{1}=1}^{M}(X_{ij}-\hat{\theta})(X_{ij_{1}}-\hat{\theta})^{\prime}+\frac{1}{N^{2}M^{2}}\sum_{i=1}^{N}\sum_{i_{1}=1}^{N}\sum_{j=1}^{M}(X_{ij}-\hat{\theta})(X_{i_{1}j}-\hat{\theta})^{\prime}\\
 &  & -\frac{1}{N^{2}M^{2}}\sum_{i=1}^{N}\sum_{j=1}^{M}(X_{ij}-\hat{\theta})(X_{ij}-\hat{\theta})^{\prime}.
\end{eqnarray*}
The variance estimator $\hat{\Sigma}$ is a two-way version of the cluster robust variance estimator of \citet{liang1986longitudinal} from \citet{miglioretti2007marginal}, \citet{CGM2011}, and \citet{thompson2011simple}. Its asymptotic properties are subsequently investigated in \citet{davezies2021} and \citet{mackinnon2021wild}.  In terms of the first-order asymptotic property, both $\ell^{m}(\theta)$ and $T(\theta)$ converge in distribution to the $\chi_{d}^{2}$ distribution. 

Denote $\Phi$ and $\phi$ be the standard normal cumulative distribution and density functions, respectively, and $\mathrm{i}=\sqrt{-1}$. Higher-order properties of $\ell^{m}(\theta)$ and $T(\theta)$ under the simplified setup in (\ref{eq:simple}) are presented as follows.

\begin{thm} \label{thm:high} 
{  Consider the setup in (\ref{eq:simple}), and assume  $E[\varepsilon_{ij}^{10}]<\infty$ and the Cramér condition $\underset{|t|\to\infty}{\lim\sup}|E[e^{\mathrm{i}t\varepsilon_{ij}}]|<1$. Furthermore suppose $Var(\varepsilon_{ij})=1$ to simplify the presentation.} Then for each $t>0$,
\[
\Pr\left\{\sqrt{T(\theta)}\le t\right\} = \Phi(t)-\left\{ \frac{3}{2}\left(\frac{1}{N}+\frac{1}{M}\right)t+\frac{1}{2}\left(\frac{1}{N}+\frac{1}{M}\right)t^{3}\right\} \phi(t)+o(n^{-1}),
\]
and
\begin{eqnarray*}
\Pr\left\{\sqrt{\ell^{m}(\theta)}\le t\right\} & = & \Phi(t)-\left\{ \left( \frac{3}{2}\left(\frac{1}{N}+\frac{1}{M}\right)-\frac{3}{N}-\frac{3}{M}+\frac{5}{n}\right) t+\frac{1}{2}\left(\frac{1}{N}+\frac{1}{M}-\frac{2(\sqrt{2}-1)}{n}\right)t^{3}\right\} \phi(t)\\
 &  & +o(n^{-1}).
\end{eqnarray*}
\end{thm}

Several remarks follow. First, the asymptotic expansion for the (signed root of) Wald statistic $T(\theta)$ based on the Eicker-White type cluster robust variance estimator shows that its second-order term is of order $O(n^{-1})$ and takes a negative value. This result suggests undercoverage of the Wald-type confidence interval based on $T(\theta)$ in finite samples as illustrated in our simulation studies in Section \ref{sec:sim}.

Second, the asymptotic expansion for the modified MEL statistic $\ell^{m}(\theta)$ shows that the second-order term is also of order $O(n^{-1})$ but is closer to zero. This result indicates desirable accuracy of the modified MEL statistic by the $\chi^{2}$ calibration for the degenerate case.

Finally, even without the factor $\mathcal{C}(N,M)$ in the adjustment term in (\ref{eq:adjQ}), the corresponding modified MEL (say, $\tilde{\ell}^{m}(\theta)$) yields a smaller second-order term than that of the Wald statistic as 
\begin{eqnarray*}
\Pr\left\{\sqrt{\tilde{\ell}^{m}(\theta)}\le t\right\} & = & \Phi(t)-\left\{ \left( \frac{3}{2}\left(\frac{1}{N}+\frac{1}{M}\right)-\frac{1}{N}-\frac{1}{M}+\frac{1}{n}\right) t+\frac{1}{2}\left(\frac{1}{N}+\frac{1}{M}-\frac{2(\sqrt{2}-1)}{n}\right)t^{3}\right\} \phi(t)\\
 &  & +o(n^{-1}).
\end{eqnarray*}
However, the refinement of $\tilde{\ell}^{m}(\theta)$ in the coefficient of $t\phi(t)$ is smaller than that of $\ell^{m}(\theta)$.



\section{Generalizations\label{sec:general}}

\subsection{Bipartite network\label{sub:network}}

In this section, we extend our (modified) MEL inference method for slope parameters in the logistic regression model for sparse bipartite network models investigated by \citet{Graham2020logit}. While the asymptotic properties of the maximum composite likelihood estimator under sparse network asymptotics have been studied  in the literature, no inference method has been proposed for this estimator.

Let $\mu$, $\{(W_{i0},A_{i0}):i=1,\ldots,N\}$, $\{(W_{0j},A_{0j}):j=1,\ldots,M\}$, and $\{V_{ij}:i=1,\ldots,N,j=1,\ldots,M\}$ be i.i.d. sequences, where $W_{i0}$ and $W_{0j}$ are observed attributes with supports $\mathcal{W}_{1}$ and $\mathcal{W}_{2}$, respectively, 
and $(\mu,A_{i0},A_{0j},V_{ij})$ are unobserved shocks. Suppose the random bipartite graph $\{Y_{ij}\in\{0,1\}:i=1,\ldots,N,j=1,\ldots,M\}$ is generated according to
\begin{equation}
Y_{ij}=h_{N,M}(\mu,W_{i0},W_{0j},A_{i0},A_{j0},V_{ij}),\label{eq:Y}
\end{equation}
where $h_{N,M}:[0,1]\times\mathcal{W}_{1}\times\mathcal{W}_{2}\times[0,1]^{3}\to\{0,1\}$ is a graphon unknown to the researcher. Suppose the researcher observes $\{Y_{ij},Z_{ij}:i=1,\ldots,N,j=1,\ldots,M\}$, where $Z_{ij}=z(W_{i0},W_{0j})$ is a vector of known transformations of $W_{i0}$ and $W_{0j}$. Consider the logistic network formation model 
\begin{equation}
\Pr(Y_{ij}=1\mid W_{i0},W_{0j})=\Lambda(\alpha_{n}+Z_{ij}^{\prime}\beta),\label{eq:logit}
\end{equation}
where $\Lambda(u)=\exp(u)/(1+\exp(u))$, and $\alpha_{n}$ is an intercept, which may vary with $n=N+M$. The asymptotics is understood as $n\to\infty$. Suppose $M/n\to\phi\in(0,1)$. Let $\rho_{n}=E[\Lambda(\alpha_{n}+Z_{11}^{\prime}\beta)]$ be the marginal link formation probability, and $\lambda_{n}^{1}=M\rho_{n}$ and $\lambda_{n}^{2}=N\rho_{n}$ be the average degrees of the first and second cluster dimensions, respectively. Suppose $\alpha_{n}=\log(\eta/n)$ for some constant $\eta$. Note that under such setting, $\alpha_{n}\to-\infty$ and $\lambda_{n}^{1}\to\lambda^{1}\in(0,\infty)$ as $n\to\infty$, that is, both average degrees stay finite in the limit.

We estimate the model by the composite maximum likelihood 
\begin{equation}
(\hat{\alpha},\hat{\beta})=\arg\max_{\alpha,\beta}\sum_{i=1}^{N}\sum_{j=1}^{M}\mathcal{L}_{ij}(\alpha,\beta),\label{eq:MLE}
\end{equation}
where $\mathcal{L}_{ij}(\alpha,\beta)=Y_{ij}\log\Lambda(\alpha_{n}+Z_{ij}^{\prime}\beta)+(1-Y_{ij})\log(1-\Lambda(\alpha_{n}+Z_{ij}^{\prime}\beta))$. Suppose the object of interest is a $d$-dimensional subvector $\theta$ of $(\alpha,\beta^{\prime})^{\prime}$. The MEL function $\ell(\theta)$ for $\theta$ is obtained as in (\ref{eq:MEL}) by setting $S(\theta)=\hat{\theta}-\theta$ and $S^{(l)}(\theta)=\hat{\theta}^{(l)}-\theta$, where $\hat{\theta}$ is the corresponding subvector of the estimator in (\ref{eq:MLE}) and $\hat{\theta}^{(l)}$ is the corresponding subvector of 
\[
(\hat{\alpha}^{(l)},\hat{\beta}^{(l)})=\left\{ \begin{array}{cc}
\arg\max_{\alpha,\beta}\sum_{i\neq l}\sum_{j=1}^{M}\mathcal{L}_{ij}(\alpha,\beta) & \text{if }i\le N,\\
\arg\max_{\alpha,\beta}\sum_{i=1}^{N}\sum_{j\neq l-N}\mathcal{L}_{ij}(\alpha,\beta) & \text{otherwise.}
\end{array}\right.
\]
Similarly, the modified MEL function $\ell^{m}(\theta)$ can be obtained as in (\ref{eq:mMEL}) by setting $S_{l,l_{1}}(\theta)=\hat{\theta}^{(l,l_{1})}-\theta$, where $\hat{\theta}^{(l,l_{1})}$ is the corresponding subvector of $\arg\max_{\alpha,\beta}\sum_{i\neq l}\sum_{j\neq l_{1}}\mathcal{L}_{ij}(\alpha,\beta)$.

The asymptotic property of $\ell^{m}(\theta)$ is obtained under the following conditions. Note that we do not need to impose a high-level condition that corresponds to Assumption \ref{asm:MEL}(iii), as it holds automatically under the current setting \citep[pp. 15]{Graham2020logit}.
\begin{asm} \label{asm:net} (i) (\ref{eq:Y}) and (\ref{eq:logit}) hold true. (ii) $(\eta,\beta^{\prime})^{\prime}$ lies in the interior of a compact parameter space. (iii) $z(\cdot,\cdot)$ is compactly supported. (iv) $M/n\to\phi\in(0,1)$ as $n\to\infty$. (v) $H=-\eta E[\exp(Z_{11}^{\prime}\beta)(1,Z_{11}^{\prime})^{\prime}(1,Z_{11}^{\prime})]$
is of full rank. \end{asm}

\begin{thm} \label{thm:net} Under Assumption \ref{asm:net}, it holds
\[
\ell^{m}(\theta)\stackrel{d}{\to}\chi_{d}^{2}.
\]
\end{thm}

Similar comments to Theorem \ref{thm:mMEL} apply. As shown in \citet{Graham2020logit}, the asymptotic variance of the maximum composite likelihood estimator $\hat{\theta}$ involves several terms due to non-negligible contributions from the higher-order terms in the Hoeffding type decomposition (\ref{eq:Hoeff}). In contrast, our modified MEL approach only requires $\hat{\theta}$, $\hat{\theta}^{(l)}$'s, and $\hat{\theta}^{(l,l_{1})}$'s, and circumvents estimation of such variance components.

\subsection{Generalized estimating equations under cluster dependence\label{sub:GEE}}

In Section \ref{sub:network}, we consider inference on a logistic regression model for bipartite network data, where the modified MEL function is constructed based on the composite maximum likelihood estimator. More generally, our MEL method can be applied to conduct inference on parameters defined via generalized estimating equations (GEEs) for longitudinal data \citep{liang1986longitudinal}.  The existing literature on the GEE mostly focuses on the case where the cluster size is fixed and there is no dependence across clusters. A notable exception is \cite{xie2003asymptotics} who investigated the asymptotic properties of the GEE estimators under the asymptotic regime of $N\to \infty$ and $M$ being either fixed or diverges to infinity at some appropriate rates while maintaining  independence across clusters. Thus it is an interesting open question whether we can conduct valid inference for parameters under both growing cluster sizes and dependence across clusters.

To fix the idea, consider a generalized linear model based on the density $f(Y_{ij}|Z_{ij},\theta,\phi)=\exp[\{Y_{ij}u(Z_{ij}^{\prime}\theta)-a(u(Z_{ij}^{\prime}\theta))+b(Y_{ij})\}/\phi]$ for $i=1,\ldots,N$ and $j=1,\ldots,M$, where $u$, $a$, and $b$ are known functions and $\phi$ is a known constant. To conduct inference on $\theta$ when $M\to\infty$ and $(Y_{ij},Z_{ij})$ is embedded into a separately exchangeable array, we employ the estimating equations using the independent working correlation matrix 
\[
\sum_{i=1}^{N}\sum_{j=1}^{M}u^{(1)}(Z_{ij}^{\prime}\theta)Z_{ij}\{Y_{ij}-a^{(1)}(u(Z_{ij}^{\prime}\theta))\}=0,
\]
where $u^{(1)}$ and $a^{(1)}$ are the derivatives of $u$ and $a$, respectively. Letting $X_{ij}(\theta)=u^{(1)}(Z_{ij}^{\prime}\theta)Z_{ij}\{Y_{ij}-a^{(1)}(u(Z_{ij}^{\prime}\theta))\}$, the modified MEL function $\ell^{m}(\theta)$ is defined as in (\ref{eq:mMEL}) by setting $S(\theta)=\frac{1}{NM}\sum_{i=1}^{N}\sum_{j=1}^{M}X_{ij}(\theta)$,
\begin{eqnarray*}
S_{l}(\theta) & = & \left\{ \begin{array}{cc}
\frac{1}{(N-1)M}\sum_{i\ne l}\sum_{j=1}^{M}X_{ij}(\theta) & \text{if }l\le N,\\
\frac{1}{N(M-1)}\sum_{i=1}^{N}\sum_{j\ne l-N}X_{ij}(\theta) & \text{otherwise,}
\end{array}\right.\\
S_{l,l_{1}}(\theta) & = & \frac{1}{(N-1)(M-1)}\sum_{i\ne l}\sum_{j\ne l_{1}}X_{ij}(\theta).
\end{eqnarray*}
Then as far as the assumptions for Theorem \ref{thm:mMEL} are satisfied for $X_{ij}(\theta)$, we obtain $\ell^{m}(\theta)\overset{d}{\to}\chi_{\dim (\theta)}^{2}$ at the true value of $\theta$. Also the modified MEL statistic for the composite null hypothesis $H_{0}:r(\theta)=0$ can be obtained by $\min_{\theta:r(\theta)=0}\ell^{m}(\theta)$, which converges to $\chi_{\dim (r(\theta))}^{2}$ by adapting the argument in \cite{qin1994empirical}. 

\subsection{Multiway MEL\label{sub:three-way}}

Let us now extend the (modified) MEL approach to the three way case $\{X_{ijt}:i=1,\ldots,N,j=1,\ldots,M,t=1,\ldots,T\}$. Similar modification works for $K$-way for any fixed $K\in\mathbb{N}$. Consider the sample mean $\hat{\theta}=(NMT)^{-1}\sum_{i=1}^{N}\sum_{j=1}^{M}\sum_{t=1}^{T}X_{ijt}$ for the population mean $\theta=E[X_{111}]$. Note that following the Aldous-Hoover representation as in (\ref{eq:AH}) and the Hoeffding type decomposition \citep[Lemma 1]{ChiangKatoSasaki2020} for general $K$-way mean, we have 
\begin{eqnarray*}
\hat{\theta}-\theta & = & \frac{1}{N}\sum_{i=1}^{N}L_{i00}+\frac{1}{M}\sum_{j=1}^{M}L_{0j0}+\frac{1}{T}\sum_{t=1}^{T}L_{00t}\\
 &  & +\frac{1}{NM}\sum_{i=1}^{N}\sum_{j=1}^{M}A_{ij0}+\frac{1}{MT}\sum_{j=1}^{M}\sum_{t=1}^{T}A_{0jt}+\frac{1}{NT}\sum_{i=1}^{N}\sum_{t=1}^{T}A_{i0t}+\frac{1}{NMT}\sum_{i=1}^{N}\sum_{j=1}^{M}\sum_{t=1}^{T}A_{ijt},
\end{eqnarray*}
where 
\begin{eqnarray*}
L_{i00} & = & E[X_{i11}\mid U_{i00}]-\theta,\quad L_{0j0}=E[X_{1j1}\mid U_{0j0}]-\theta,\quad L_{00t}=E[X_{11t}\mid U_{00t}]-\theta,\\
A_{ij0} & = & E[X_{ij1}\mid U_{i00},U_{0j0},U_{ij0}]-L_{i00}-L_{0j0}-\theta,\\
A_{i0t} & = & E[X_{i1t}\mid U_{i00},U_{00t},U_{i0t}]-L_{i00}-L_{00t}-\theta,\\
A_{0jt} & = & E[X_{1jt}\mid U_{0j0},U_{00t},U_{0jt}]-L_{0j0}-L_{00t}-\theta,\\
A_{ijt} & = & X_{ijt}-A_{ij0}-A_{i0t}-A_{0jt}-L_{i00}-L_{0j0}-L_{00t}-\theta.
\end{eqnarray*}
 Suppose it holds that
\[
\sum_{i=1}^{N}\sum_{j=1}^{M}A_{ij0}=\sum_{i=1}^{N}\sum_{j=1}^{M}R_{ij0}\{1+o_{p}(1)\},
\]
where $R_{ij0}=g_{12}(U_{ij0})$, $g_{12}:[0,1]\to\mathbb{R}^{\dim(\theta)}$ is an unknown Borel-measurable mapping. Similarly, suppose that $\sum_{i=1}^{N}\sum_{t=1}^{T}A_{i0t}=\sum_{i=1}^{N}\sum_{t=1}^{T}\{R_{i0t}+o_{p}(1)\}$ with $R_{i0t}=g_{13}(U_{i0t})$, $\sum_{j=1}^{M}\sum_{t=1}^{T}A_{0jt}=\sum_{j=1}^{M}\sum_{t=1}^{T}\{R_{0jt}+o_{p}(1)\}$ with $R_{0jt}=g_{23}(U_{0jt})$, and $\sum_{i=1}^{N}\sum_{j=1}^{M}\sum_{t=1}^{T}A_{ijt}=\sum_{i=1}^{N}\sum_{j=1}^{M}\sum_{t=1}^{T}\{R_{ijt}+o_{p}(1)\}$ with $R_{ijt}=g_{123}(U_{ijt})$ for some unknown Borel-measurable mappings $g_{13}$, $g_{23}$, and $g_{123}$. This is analogous to Assumption \ref{asm:MEL}(iii). Only in this subsection, let $n=N+M+T$. Then a central limit theorem implies that as $\min\{N,M,T\}\to\infty$, we have $\sqrt{n}(\hat{\theta}-\theta)\overset{d}{\to}N(0,\sigma_{*}^{2})$, where 
\[
\sigma_{*}^{2}=n\left\{ \frac{Var(L_{100})}{N}+\frac{Var(L_{010})}{M}+\frac{Var(L_{001})}{T}+\frac{Var(R_{110})}{NM}+\frac{Var(R_{101})}{NT}+\frac{Var(R_{011})}{MT}+\frac{Var(R_{111})}{NMT}\right\}.
\]

Now, define $S(\theta)=\hat{\theta}-\theta$, $S_{l}(\theta)=\hat{\theta}^{(l)}-\theta$, and the leave-one-index-out estimators
\[
\hat{\theta}^{(l)}=\left\{ \begin{array}{cc}
\frac{1}{(N-1)MT}\sum_{i\ne l}\sum_{j=1}^{M}\sum_{t=1}^{T}X_{ijt} & \text{ if \ensuremath{l\le N}},\\
\frac{1}{N(M-1)T}\sum_{i=1}^{N}\sum_{j\ne l+N}\sum_{t=1}^{T}X_{ijt} & \text{ if \ensuremath{N<l\le N+M}},\\
\frac{1}{NM(T-1)}\sum_{i=1}^{N}\sum_{j=1}^{M}\sum_{t\ne l+N+M}^{T}X_{ijt} & \text{ otherwise}.
\end{array}\right.
\]
Further, define $V_{l}(\theta)=nS(\theta)-(n-1)S^{(l)}(\theta)$, $V_{l}^{m}(\theta)=V_{l}(\hat{\theta})-\hat{\Gamma}\tilde{\Gamma}^{-1}\{V_{l}(\hat{\theta})-V_{l}(\theta)\},$ for $l=1,\ldots,N$, where $\hat{\Gamma}$ and $\tilde{\Gamma}$ are so that 
\begin{eqnarray*}
\hat{\Gamma}\hat{\Gamma}^{\prime} & = & \frac{1}{n}\sum_{l=1}^{n}V_{l}(\hat{\theta})V_{l}(\hat{\theta})^{\prime},\\
\tilde{\Gamma}\tilde{\Gamma}^{\prime} & = & \frac{1}{n}\sum_{l=1}^{n}V_{l}(\hat{\theta})V_{l}(\hat{\theta})^{\prime}-\frac{1}{n}\left\{ \sum_{l=1}^{N}\sum_{l_{1}=1}^{M}Q_{ll_{1}0}Q_{ll_{1}0}^{\prime}+\sum_{l_{1}=1}^{M}\sum_{l_{2}=1}^{T}Q_{0l_{1}l_{2}}Q_{0l_{1}l_{2}}^{\prime}+\sum_{l=1}^{N}\sum_{l_{2}=1}^{T}Q_{l0l_{2}}Q_{l0l_{2}}^{\prime}\right\} \\
 &  & +\frac{1}{n}\sum_{l=1}^{N}\sum_{l_{1}=1}^{M}\sum_{l_{2}=1}^{T}Q_{ll_{1}l_{2}}Q_{ll_{1}l_{2}}^{\prime},
\end{eqnarray*}
where 
\begin{eqnarray*}
	Q_{ll_{1}0} & = &  [n\hat{\theta}-(n-1)(\hat{\theta}^{(l)}+\hat{\theta}^{(N+l_{1})})+(n-2)\hat{\theta}^{(l,l_{1},0)}],\\
	Q_{l0l_{2}} & = & [n\hat{\theta}-(n-1)(\hat{\theta}^{(l)}+\hat{\theta}^{(N+M+l_{2})})+(n-2)\hat{\theta}^{(l,0,l_{2})}],\\
	Q_{0l_{1}l_{2}} & = &[ n\hat{\theta}-(n-1)(\hat{\theta}^{(N+l_{1})}+\hat{\theta}^{(N+M+l_{2})})+(n-2)\hat{\theta}^{(0,l_{1},l_{2})}],\\
	Q_{ll_{1}l_{2}} & = & [n\hat{\theta}-(n-1)(\hat{\theta}^{(l)}+\hat{\theta}^{(N+l_{1})}+\hat{\theta}^{(N+M+l_{2})})\\
	&  & +(n-2)(\hat{\theta}^{(l,l_{1},0)}+\hat{\theta}^{(0,l_{1},l_{2})}+\hat{\theta}^{(l,0,l_{2})})-(n-3)\hat{\theta}^{(l,l_{1},l_{2})}],
\end{eqnarray*}
and the leave-two-index-out and leave-three-index-out estimators are defined by 
\begin{eqnarray*}
\hat{\theta}^{(l,l_{1},0)} & = & \frac{1}{(N-1)(M-1)T}\sum_{i\ne l}\sum_{j\ne l_{1}}\sum_{t=1}^{T}X_{ijt}\quad\text{for }\ensuremath{l\le N},\ensuremath{l_{1}\le M},\\
\hat{\theta}^{(0,l_1,l_{2})} & = & \frac{1}{N(M-1)(T-1)}\sum_{i=1}^{N}\sum_{j\ne l_1}\sum_{t\ne l_{2}}X_{ijt}\quad\text{for }\ensuremath{l_1\le M},\ensuremath{l_{2}\le T},\\
\hat{\theta}^{(l,0,l_{2})} & = & \frac{1}{(N-1)M(T-1)}\sum_{i\ne l}\sum_{j=1}^{M}\sum_{t\ne l_{2}}X_{ijt}\quad\text{for }\ensuremath{l\le N},\ensuremath{l_{2}\le T},\\
\hat{\theta}^{(l,l_{1},l_{2})} & = & \frac{1}{(N-1)(M-1)(T-1)}\sum_{i\ne l}\sum_{j\ne l_{1}}\sum_{t\ne l_{2}}X_{ijt}\quad\text{for }\ensuremath{l\le N},\ensuremath{l_{1}\le M},\ensuremath{l_{2}\le T}.
\end{eqnarray*}
Under some regularity conditions, it can be shown similarly as in the proof of Theorem \ref{thm:MEL} that 
\begin{eqnarray*}
\frac{1}{n}\sum_{l=1}^{n}V_{l}(\theta)V_{l}(\theta)^{\prime} & = & n\Bigg\{\Bigg(\frac{Var(L_{100})}{N}+\frac{Var(L_{010})}{M}+\frac{Var(L_{001})}{T}\Bigg)\\
 &  & \quad+2\Bigg(\frac{Var(R_{110})}{NM}+\frac{Var(R_{101})}{NT}+\frac{Var(R_{011})}{MT}\Bigg)+3\Bigg(\frac{Var(R_{111})}{NMT}\Bigg)\Bigg\}+o_{p}(1).
\end{eqnarray*}
We can then obtain $\ell^{m}(\theta)$ by following the same definition as in (\ref{eq:mMEL}) with corresponding components replaced by those defined in this section. It can be shown that under regularity conditions, the modified MEL statistic has the pivotal asymptotic distribution 
\[
\ell^{m}(\theta)\stackrel{d}{\to}\chi_{\dim(\theta)}^{2}.
\]
In fact, this proposed procedure is much less computationally intensive in comparison with the corresponding jackknife procedures for $U$-statistics since the number of all leave-out estimators in the proposed procedure is $O(n^3)$, the same as the order of all different leave-one-out estimators for i.i.d. data. On the other hand, the conventional leave-three-out estimators with sample size $NMT$ consist of ${NMT \choose 3}=O(n^9)$ possibilities.

\section{Simulation\label{sec:sim}}

This section conducts a simulation study to evaluate the finite sample properties of the proposed MEL inference methods. In particular, we consider a random effect model (Section \ref{sub:sima}) and bipartite stochastic block model (Section \ref{sub:simb}). We shall focus on simple means based on the reasoning as in \citet{Owen2007} that one can expect a method that gives the correct variance for a mean to be reliable for more complicated statistics such as smooth functions of means and estimating equation parameters.

\subsection{Random effect model\label{sub:sima}}

We first consider the random effect model studied in \citet{Owen2007} and \citet[Ch 5]{searle2009variance}:
\[
X_{ij}=\theta+a_{i}+b_{j}+\varepsilon_{ij},
\]
where $\theta=1$ and $(a_{i},b_{j},\varepsilon_{ij})$ are mutually independent random variables with $a_{i},b_{j}\sim N(0,\sigma^{2})$ and $\varepsilon_{ij}\sim N(0,1)$. The estimator considered here is the sample mean. We vary $\sigma^{2}\in\{1,0.1,0\}$ to examine the performance under non-degenerate, nearly degenerate, and degenerate cases, respectively. We set $N=50$ and $M\in\{5,10,15,20,30,50\}$.

We compare seven methods of constructing confidence intervals: (i) multiway empirical likelihood (MEL), (ii) modified MEL (mMEL), (iii) Wald confidence interval with modified multiway variance estimator from Remark \ref{rem:mMW} (mMW), (iv) Wald with Eicker-White type multiway cluster robust variance estimator (EWW), (v) bootstrap with model selection  (MBS), (vi) conservative bootstrap (MBC), and (vii) Wald with i.i.d. variance estimator (IID). The methods (i)-(iii) are our developments, (iv) is a conventional method, (v)-(vi) are proposed by \cite{menzel2021bootstrap}, and (vii) is asymptotically invalid (except for the degenerate case) but included for comparison. The nominal coverage is set as $0.95$.

Table 1 reports empirical coverages of the methods (i)-(vii) based on $5,000$ Monte Carlo replications. Our findings are summarized as follows. First, IID does not work at all except for the degenerate case (i.e., $\sigma^{2}=0$). Since IID is asymptotically invalid, its size distortion remains even for $N,M=50$. Second, EWW exhibits severe under-coverages when $M$ is small, as predicted by the higher-order analysis in Section \ref{sub:2nd}. Third, mMEL, which is asymptotically valid for all cases, outperforms in almost all cases. Even if $M$ is small, mMEL performs well. Fourth, MEL works well for non-degenerate case (i.e., $\sigma^{2}=1$) but over-covers for nearly degenerate and degenerate cases. This result is expected from Theorem \ref{thm:MEL}. As shown in Theorem \ref{thm:mMEL}, mMEL recovers asymptotic pivotalness for all cases and our simulation result clearly illustrates this point. Fifth, for Wald-type confidence  intervals, the proposed mMW works better than the conventional EWW but slightly under-covers. Finally, both MBS and MBC show significant over-coverages.
Overall we recommend mMEL, which exhibits accurate coverages and is robust for all cases.

\begin{table}
	\begin{tabular}{c|c|ccccccc}
		\hline 
		$M$ & $\theta$ & MEL & mMEL & mMW & EWW & MBS&MBC&IID\tabularnewline
		\hline
		& $1$  & 0.940&	0.938&		0.918&		0.860&	0.998&0.998&	0.324\tabularnewline
		$5$  & $0.1$  & 0.960&0.941&0.923&0.861&0.994&0.994&0.604 \tabularnewline
		& $0$  &  0.991 &0.941&0.934&	0.820	&0.986&0.986&	0.946 \tabularnewline
		
		\hline 
		& $1$  & 0.954&0.952&0.935&0.916&1.000&1.000&0.323\tabularnewline
		$10$  & $0.1$ & 0.962&0.946&0.935&0.908&1.000&1.000&0.586 \tabularnewline
		& $0$  & 0.992&0.943&0.937&0.890&0.997&0.997&0.946\tabularnewline
		\hline 
		& $1$  &0.953&0.953&0.951&0.939&1.000&1.000&0.315 \tabularnewline
		$15$  & $0.1$  & 0.965&0.952&0.943&0.931&1.000&1.000&0.560\tabularnewline
		& $0$  &  0.994&0.939&0.934&0.912&1.000&1.000&0.946\tabularnewline
		\hline 
		& $1$  & 0.950&0.948&0.940&0.930&1.000&1.000&0.314\tabularnewline
		$20$  & $0.1$  & 0.965&0.954&0.948&0.934&1.000&1.000&0.552\tabularnewline
		& $0$  &0.993&0.941&0.938&0.913&1.000&1.000&0.950 \tabularnewline
		\hline 
		& $1$  &0.948&0.945&0.944&0.940&1.000&1.000&0.296 \tabularnewline
		$30$  & $0.1$  &0.960&0.949&0.946&0.938&1.000&1.000&0.513 \tabularnewline
		& $0$  & 0.994&0.941&0.937&0.924&1.000&1.000&0.946 \tabularnewline
		\hline 
		& $1$  & 0.949&0.947&0.946&0.942&1.000&1.000&0.261\tabularnewline
		$50$  & $0.1$  &0.956&0.948&0.947&0.942&1.000&1.000&0.470 \tabularnewline
		& $0$  & 0.993&0.941&0.939&0.930&1.000&1.000&0.950\tabularnewline
		\hline 
	\end{tabular}
	
	\caption{Coverage rates for random effect model with $N=50$}
\end{table}

\subsection{Bipartite stochastic block model\label{sub:simb}}

We next consider a stochastic block model, which is an adapted version of \citet[Sec. 5.2]{bhattacharyya2015subsampling} for bipartite graphs with two distinctive community dimensions. First, each $i$ is randomly assigned to a membership $a\in\{1,2\}$ of the first community dimension with probabilities $\pi_{1}=(0.7,0.3)^{\prime}$ and each $j$ is randomly assigned to a membership $b\in\{1,2\}$ of the second community dimension with probabilities $\pi_{2}=(0.2,0.8)^{\prime}$. Then consider the following edge formation probabilities
\[
F_{ab}=\Pr(X_{ij}=1|i\in A_a,j\in B_b)=s_{\theta}S_{ab},\text{ for }a\in\{1,2\}\text{ and }b\in\{1,2\},
\]
where the blocks $A_a$ and $B_b$ satisfy $A_1\cup A_2=\{1,...,N\}$, $A_1\cap A_2=\emptyset$, $B_1\cup B_2=\{1,...,M\}$, $B_1\cap B_2=\emptyset$,  $S_{ab}$'s are elements of $S=\begin{bmatrix}0.6 & 0.4\\
0.3 & 0.7
\end{bmatrix}$, and $s_{\theta}$ is chosen to satisfy $\theta=\pi_{1}^{\prime}F\pi_{2}\in\{0.5,0.1,0.05\}$. 

Similar to the last subsection, we consider the seven confidence intervals (i)-(vii), as introduced in the previous subsection, for $\theta$ with the nominal coverage $0.95$ for the cases of $N=50$ and $M\in\{5,10,15,20,30,50\}$. Table 2 reports empirical coverages of the methods (i)-(vii) based on $5,000$ Monte Carlo replications. The results are qualitatively similar to the ones in the last subsection. IID has size distortions, EWW exhibits under-coverages when $M$ is small, both MBS and MBC  over-covering across dense and sparse cases, and mMEL outperforms the rest for almost all cases. It is worthy of pointing out that in this setup, MEL shows over-coverages for all cases, including the relatively dense case (i.e., $\theta=0.5$). Therefore, we recommend to use mMEL for this simulation study, too.

\begin{table}
	\begin{tabular}{c|c|ccccccc}
		\hline 
		$M$ & $\theta$ & MEL & mMEL & mMW & EWW & MBS&MBC&IID\tabularnewline
		\hline 
		& $0.5$ &0.989&0.942&0.936&0.841&0.968&0.968&0.933\tabularnewline
		$5$ & $0.1$ &0.989&0.936&0.925&0.814&0.970&0.989&0.928 \tabularnewline
		& $0.05$ &0.988&0.939&0.920&0.831&0.993&0.999&0.924\tabularnewline
		\hline 
		& $0.5$ &0.988&0.944&0.937&0.900&0.995&0.995&0.906\tabularnewline
		$10$ & $0.1$ &0.991&0.941&0.933&0.888&0.997&0.997&0.939\tabularnewline
		& $0.05$ & 0.989&0.936&0.928&0.887&0.988&1.000&0.926\tabularnewline
		\hline 
		& $0.5$ &0.990&0.945&0.939&0.920&0.999&0.999&0.903\tabularnewline
		$15$ & $0.1$ &0.992&0.945&0.937&0.914&0.996&1.000&0.940\tabularnewline
		& $0.05$ &0.992&0.940&0.931&0.905&0.998&1.000&0.944\tabularnewline
		\hline 
		& $0.5$ &0.986&0.947&0.942&0.926&0.999&0.999&0.874
		\tabularnewline
		$20$ & $0.1$ &0.990&0.934&0.930&0.909&1.000&1.000&0.937\tabularnewline
		& $0.05$ &0.993&0.938&0.931&0.909&0.997&1.000&0.935 \tabularnewline
		\hline 
		& $0.5$ &0.983&0.943&0.940&0.929&1.000&1.000&0.848\tabularnewline
		$30$ & $0.1$ &0.991&0.944&0.941&0.930&1.000&1.000&0.941\tabularnewline
		& $0.05$ & 0.991&0.942&0.938&0.923&0.999&1.000&0.939\tabularnewline
		\hline 
		& $0.5$ &0.981&0.948&0.945&0.939&1.000&1.000&0.788\tabularnewline
		$50$ & $0.1$ &0.991&0.948&0.945&0.936&1.000&1.000&0.931\tabularnewline
		& $0.05$ & 0.994&0.948&0.944&0.934&1.000&1.000&0.943\tabularnewline
		\hline 
	\end{tabular} \caption{Coverage rates for stochastic block model with $N=50$}
\end{table}
\newpage{}

\appendix

\section{Mathematical appendix}

\subsection{Proof of Theorem \ref{thm:MEL}}

In this subsection, we use the following lemma.

\begin{lem}\label{lem:maxVl} Suppose that $E[\|X_{11}\|^{q}]\le C<\infty$ and $\rho=n/\underline{n}=o(n^{1/2-2/q})$ for some $q>4$. Then it holds $\max_{1\le l\le n}\|V_{l}(\theta)\|=o_{p}(n^{1/2})$. \end{lem}

\subsubsection*{Proof of Lemma \ref{lem:maxVl}.}

For notational simplicity, assume that $X_{11}$ is a scalar. For each $l=1,\ldots,N$, it holds 
\[
S(\theta)=\frac{1}{NM}\sum_{i=1}^{N}\sum_{j=1}^{M}X_{ij}-\theta=\frac{1}{NM}\sum_{j=1}^{M}(X_{lj}-\theta)+\frac{N-1}{N}S_{l}(\theta),
\]
and thus $V_{l}(\theta)=nS(\theta)-(n-1)S_{l}(\theta)$ satisfies
\begin{eqnarray*}
\max_{1\le l\le N}|V_{l}(\theta)| & = & \max_{1\le l\le N}\left|\frac{n}{NM}\sum_{j=1}^{M}(X_{lj}-\theta)+\left(n\frac{N-1}{N}-(n-1)\right)S_{l}(\theta)\right|\\
 & = & O_{p}\left(\frac{n}{N}\max_{1\le i\le N}\max_{1\le j\le N}|X_{ij}|\right)+O_{p}\left(\max_{1\le l\le N}|S_{l}(\theta)|\right)=O_{p}\left(\rho\max_{1\le i\le N}\max_{1\le j\le N}|X_{ij}|\right).
\end{eqnarray*}
Similarly, it holds that $\max_{N+1\le l\le n}|V_{l}(\theta)|=O_{p}\left(\rho\max_{1\le i\le N}\max_{1\le j\le N}|X_{ij} |\right).$ Now, by Jensen's inequality, 
\[
E\left[\max_{1\le i\le N}\max_{1\le j\le N}|X_{ij}|\right]\le\left(E\left[\max_{1\le i\le N}\max_{1\le j\le N}|X_{ij}|^{q}\right]\right)^{1/q}\le C^{1/q}(NM)^{1/q}\le C^{1/q}n^{2/q}.
\]
Combining these results, we obtain 
\[
\max_{1\le l\le n}|V_{l}(\theta)|=O_{p}(\rho n^{2/q})=o_{p}(n^{1/2}).
\]

\subsubsection*{Proof of Theorem \ref{thm:MEL}}

For notational simplicity, assume that $X_{11}$ is a scalar, as the general result follows analogously with an application of a multivariate central limit theorem. In this proof, we will frequently utilize the $U$-statistic form of the sample variance: for $\bar{x}=n^{-1}\sum_{l=1}^{n}x_{l}$
\begin{equation}
\frac{1}{n}\sum_{l=1}^{n}(x_{l}-\bar{x})^{2}=\frac{1}{n^{2}}\sum_{1\le l<l_{1}\le n}(x_{l}-x_{l_{1}})^{2}.\label{pf:var-form}
\end{equation}
We will also make use of the algebraic identity $S(\theta)=n^{-1}\sum_{l=1}^{n}S_{l}(\theta)=n^{-1}\sum_{l=1}^{n}V_{l}(\theta)$. The rest of this proof is divided into three steps.\\

\noindent \textbf{Step 1.} In this step, we shall establish the main asymptotic result for the MEL statistic. Let $\hat{\lambda}=\text{arg}\max_{\lambda}\sum_{l=1}^{n}\log(1+\lambda V_{l}(\theta))$. The first-order condition for $\hat{\lambda}$ and the fact that $(1+x)^{-1}=1-x+x^{2}(1+x)^{-1}$ give 
\[
0=\frac{1}{n}\suml\frac{V_{l}(\theta)}{1+\hat{\lambda}V_{l}(\theta)}=\frac{1}{n}\suml V_{l}(\theta)-\frac{1}{n}\suml V_{l}(\theta)^{2}\hat{\lambda}+\frac{1}{n}\suml\frac{V_{l}(\theta)^{3}\hat{\lambda}^{2}}{1+\hat{\lambda}V_{l}(\theta)}.
\]
Since Lemma \ref{lem:maxVl} guarantees $\max_{1\le l\le n}|V_{l}(\theta)|=o_{p}(n^{1/2})$, a standard argument as in \citep[eq. (2.14)]{Owen1990} yields $\hat{\lambda}=O_{p}(n^{-1/2})$. Thus, one has 
\[
\hat{\lambda}=\frac{\suml V_{l}(\theta)}{\suml V_{l}(\theta)^{2}}+o_{p}(n^{-1/2}).
\]
A Taylor expansion yields 
\begin{eqnarray}
\ell(\theta) & = & 2\suml\log(1+\hat{\lambda}V_{l}(\theta))=2\suml\left(\hat{\lambda}V_{l}(\theta)-\frac{1}{2}\{\hat{\lambda}V_{l}(\theta)\}^{2}\right)+o_{p}(1)\nonumber \\
 & = & \frac{\left\{ n^{-1/2}\suml V_{l}(\theta)\right\} ^{2}}{n^{-1}\suml V_{l}(\theta)^{2}}+o_{p}(1).\label{pf:asy-exp}
\end{eqnarray}
Therefore, the conclusion follows from the convergence in distribution of the numerator and the consistency of the denominator for the non-degenerate case (Step 2) and nearly degenerate case (Step 3).\\

\noindent \textbf{Step 2.} We shall now investigate the limiting behaviors of the numerator and denominator in (\ref{pf:asy-exp}) under non-degeneracy. Under this scenario, it holds that $\underline{n}\sigma_{L}^{2}\to\infty$, which implies $\sigma_{R}^{2}=o(\underline{n}\sigma_{L}^{2})$. Also note that $\lim_{n\to\infty}\sigma_{L}^{2}/n<\infty$ under Assumption \ref{asm:MEL}(ii). Define $\sigma^{2}=\sigma_{L}^{2}/n+\sigma_{R}^{2}/(NM)$. In this case, it is sufficient for the conclusion of the theorem to show that 
\begin{eqnarray}
\frac{1}{n\sigma}\sum_{l=1}^{n}V_{l}(\theta) & \stackrel{d}{\to} & N(0,1),\label{pf:step2-1}\\
\frac{1}{n^{2}\sigma^{2}}\sum_{l=1}^{n}V_{l}(\theta)^{2} & \stackrel{p}{\to} & 1.\label{pf:step2-2}
\end{eqnarray}
We first show (\ref{pf:step2-1}). Eq. (\ref{eq:Hoeff}) and Assumption \ref{asm:MEL}(iii) imply 
\[
\frac{1}{n}\sum_{l=1}^{n}V_{l}(\theta)=\{1+o_{p}(1)\}\left\{ \frac{1}{N}\sum_{i=1}^{N}L_{i0}+\frac{1}{M}\sum_{j=1}^{M}L_{0j}\right\} +\frac{1}{NM}\sum_{i=1}^{N}\sum_{j=1}^{M}R_{ij}.
\]
For each $n$, define the sequence $\{Y_{n,t}:t=1,\ldots,n+NM\}$ by 
\begin{equation}
Y_{n,t}=\begin{cases}
L_{t0}/N & \text{for }t\le N,\\
L_{0(t-N)}/M & \text{for }N+1\le t\le n,\\
R_{ij}/NM & \text{for }\ensuremath{t>n},\text{ where }j=\lceil(t-n)/N\rceil,i=(t-n)-(j-1)N.
\end{cases}\label{pf:MDS}
\end{equation}
Notice that $n^{-1}\sum_{l=1}^{n}V_{l}(\theta)=\sum_{t=1}^{n+NM}Y_{n,t}+o_{p}(1)$. It is straightforward to verify that it has zero-mean and for any $t\ge 1$, $E[Y_{n,t}|\{Y_{n,t_{1}}:t_{1}=1,\ldots,t-1\}]=0$, i.e., $\{Y_{n,t}:t=1,\ldots,n+NM\}$ is a martingale difference sequence. Moreover, observe that 
\[
\sum_{t=1}^{n+NM}E[(Y_{n,t}/\sigma)^{2+\delta}]\to0\text{ for some \ensuremath{\delta>0}},\text{ and }\sum_{t=1}^{n+NM}Y_{n,t}^{2}-\sigma^{2}\stackrel{p}{\to}0,
\]
where the first result (Lyapunov's condition) is implied by Assumption \ref{asm:MEL}(ii) and the condition that $\underline{n}\sigma_{L}^{2}\to\infty$, and the second result follows from the weak law of large numbers and conditional independence of $\{R_{ij}:i=1,\ldots,N,j=1,\ldots,M\}$. Therefore, the central limit theorem for martingale difference triangular arrays \citep[Corollary 5.26]{white2001asymptotic} implies 
\[
\sum_{t=1}^{n+NM}\frac{Y_{n,t}}{\sigma}=\frac{1}{n\sigma}\sum_{l=1}^{n}V_{l}(\theta)+o_{p}(1)\stackrel{d}{\to}N(0,1),
\]
i.e., the result in (\ref{pf:step2-1}) follows.

We now show (\ref{pf:step2-2}). Note that 
\begin{eqnarray*}
 &  & \frac{1}{n^{2}\sigma^{2}}\sum_{l=1}^{n}V_{l}(\theta)^{2}=\frac{1}{n^{2}\sigma^{2}}\sum_{l=1}^{n}[S(\theta)-(n-1)\{S_{l}(\theta)-S(\theta)\}]^{2}\\
 & = & \frac{1}{n\sigma^{2}}S(\theta)^{2}+\frac{(n-1)^{2}}{n^{2}\sigma^{2}}\sum_{l=1}^{n}\{S_{l}(\theta)-S(\theta)\}^{2}=\frac{(n-1)^{2}}{n^{2}\sigma^{2}}\sum_{l=1}^{n}\{S_{l}(\theta)-S(\theta)\}^{2}+o_{p}(1)\\
 & = & \frac{(n-1)^{2}}{n^{2}\sigma^{2}}\left[\sum_{i=1}^{N}\left\{ S_{i}(\theta)-\frac{1}{N}\sum_{i=1}^{N}S_{i}(\theta)\right\} ^{2}+\sum_{j=1}^{M}\left\{ S_{N+j}(\theta)-\frac{1}{M}\sum_{j=1}^{M}S_{N+j}(\theta)\right\} ^{2}\right]+o_{p}(1)\\
 & = & \frac{(n-1)^{2}}{n^{2}\sigma^{2}}\left[\frac{1}{N}\sum_{i=1}^{N}\sum_{i^{\prime}=i+1}^{N}\{S_{i}(\theta)-S_{i^{\prime}}(\theta)\}^{2}+\frac{1}{M}\sum_{j=1}^{M}\sum_{j^{\prime}=j+1}^{M}\{S_{N+j}(\theta)-S_{N+j^{\prime}}(\theta)\}^{2}\right]+o_{p}(1)\\
 & = & \frac{(n-1)^{2}}{n^{2}N\sigma^{2}}\sum_{i=1}^{N}\sum_{i^{\prime}=i+1}^{N}\left[\frac{1}{N-1}(L_{i^{\prime}0}-L_{i0})+\frac{1}{(N-1)M}\sum_{l=1}^{M}\{(W_{i^{\prime}l}-W_{il})+(R_{i^{\prime}l}-R_{il})\}\right]^{2}\\
 &  & +\frac{(n-1)^{2}}{n^{2}M\sigma^{2}}\sum_{j=1}^{M}\sum_{j^{\prime}=j+1}^{M}\left[\frac{1}{M-1}(L_{0j^{\prime}}-L_{0j})+\frac{1}{N(M-1)}\sum_{k=1}^{N}\{(W_{kj^{\prime}}-W_{kj})+(R_{kj^{\prime}}-R_{kj})\}\right]^{2}+o_{p}(1)\\
 & = & \frac{\sigma_{L}^{2}}{n\sigma^{2}}+o_{p}(1)=1+o_{p}(1),
\end{eqnarray*}
where the second equality follows from the identity $n^{-1}\sum_{l=1}^{n}S_{l}(\theta)=S(\theta)$, the third equality follows from $S(\theta)=O_{p}(\underline{n}^{-1/2})$, the fourth equality follows from $\frac{1}{N}\sum_{i=1}^{N}S_{i}(\theta)=\frac{1}{M}\sum_{j=1}^{M}S_{N+j}(\theta)=S(\theta)$, the fifth equality follows from (\ref{pf:var-form}), the seventh equality follows from the fact that each of $(L_{i0})_{i}$, $(L_{0j})_{j}$, $(W_{ij})_{ij}$, $(R_{ij})_{ij}$ is i.i.d. and centred, as well as an application of the law of large numbers, and the last equality follows from the fact $n\sigma^{2}=\sigma_{L}^{2}+o(1)$ when $\underline{n}\sigma_{L}^{2}\to\infty$.\\

\noindent \textbf{Step 3.} We shall now investigate the limiting behaviours of (\ref{pf:asy-exp}) under the nearly degenerate scenario. In this case, it holds that $\sigma_{L}^{2}=O(1/\underline{n})$, which implies $\sigma_{L}^{2}/n=O(\sigma_{R}^{2}/(NM))$. Notice that  under Assumption \ref{asm:MEL}(iii) and (\ref{eq:Hoeff}), we have
\[
\frac{1}{n}\sum_{l=1}^{n}V_{l}(\theta)=\frac{1}{N}\sum_{i=1}^{N}L_{i0}+\frac{1}{M}\sum_{j=1}^{M}L_{0j}+\frac{1}{NM}\sum_{i=1}^{N}\sum_{j=1}^{M}R_{ij}(1+o_{p}(1)).
\]
For each $n$, we define the sequence $\{Y_{n,t}:1\le t\le n+NM\}$ as in (\ref{pf:MDS}) again, which is a martingale difference sequence. Using the same argument as in Step 2, the central limit theorem for martingale difference triangular arrays yields the same result in (\ref{pf:step2-1}).

\noindent Thus, it remains to show 
\begin{equation}
\frac{1}{n^{2}\sigma^{2}}\sum_{l=1}^{n}V_{l}(\theta)^{2}\stackrel{p}{\to}\lim_{n\to\infty}\frac{\sigma_{L}^{2}/n+2\sigma_{R}^{2}/(NM)}{\sigma_{L}^{2}/n+\sigma_{R}^{2}/(NM)}.\label{pf:step3-1}
\end{equation}
To this end, analogous arguments as in Step 2 (using the fact that $S(\theta)=O_{p}((NM)^{-1/2})$) yield 
\begin{eqnarray}
 &  & \frac{1}{n^{2}\sigma^{2}}\sum_{i=1}^{n}V_{i}(\theta)^{2}=\frac{1}{n^{2}\sigma^{2}}\sum_{i=1}^{n}[S(\theta)-(n-1)\{S_{i}(\theta)-S(\theta)\}]^{2}\nonumber\\
 & = & \frac{1}{n\sigma^{2}}S(\theta)^{2}+\frac{(n-1)^{2}}{n^{2}\sigma^{2}}\sum_{i=1}^{n}\{S_{i}(\theta)-S(\theta)\}^{2}=\frac{(n-1)^{2}}{n^{2}\sigma^{2}}\sum_{i=1}^{n}\{S_{i}(\theta)-S(\theta)\}^{2}+o_{p}(1)\nonumber\\
 & = & \frac{(n-1)^{2}}{n^{2}\sigma^{2}}\left[\frac{1}{N}\sum_{i=1}^{N}\sum_{i^{\prime}=i+1}^{N}\{S_{i}(\theta)-S_{i^{\prime}}(\theta)\}^{2}+\frac{1}{M}\sum_{j=1}^{M}\sum_{j^{\prime}=j+1}^{M}\{S_{N+j}(\theta)-S_{N+j^{\prime}}(\theta)\}^{2}\right]+o_{p}(1)\nonumber\\
 & = & \frac{(n-1)^{2}}{n^{2}N\sigma^{2}}\sum_{i=1}^{N}\sum_{i^{\prime}=i+1}^{N}\left[\frac{1}{N-1}(L_{i^{\prime}0}-L_{i0})+\frac{1}{(N-1)M}\sum_{l=1}^{M}\{(W_{i^{\prime}l}-W_{il})+(R_{i^{\prime}l}-R_{il})\}\right]^{2}\nonumber\\
 & + & \frac{(n-1)^{2}}{n^{2}M\sigma^{2}}\sum_{j=1}^{M}\sum_{j^{\prime}=j+1}^{M}\left[\frac{1}{M-1}(L_{0j^{\prime}}-L_{0j})+\frac{1}{N(M-1)}\sum_{k=1}^{N}\{(W_{kj^{\prime}}-W_{kj})+(R_{kj^{\prime}}-R_{kj})\}\right]^{2}+o_{p}(1)\nonumber.
\end{eqnarray}
where the second equality follows from the identity $n^{-1}\sum_{l=1}^{n}S_{l}(\theta)=S(\theta)$, the third equality follows from $S(\theta)=O_{p}((NM)^{-1/2})$, the fourth equality follows from (\ref{pf:var-form}). { 
We now claim that RHS above equals
\begin{align*}
	\lim_{n\to \infty} \frac{\sigma_{L}^{2}/n+2\sigma_{R}^{2}/(NM)}{\sigma_{L}^{2}/n+\sigma_{R}^{2}/(NM)}+o_{p}(1).
\end{align*}
 First note that the sums involving linear components $L_{i0}$ and $L_{0j}$ can be shown to be consistent for $\lim_{n\to \infty}\sigma_{L}^2/(\sigma_{L}^2 +2n\sigma_{R}^2/(NM))$ by applying the same arguments as in the end of Step 2. For the terms involving quadratic components $R_{ij}$ and $W_{ij}$, notice that conditioning on $(U_{i0},U_{0j}:i=1,...,N,j=1,...M)$,  the terms $(R_{ij})_{ij}$ are independent over $i,j$ and satisfies $E[R_{ij}|U_{i0},U_{0j}]=0$.  It follows that
 \begin{align*}
\sigma_{R}^2=Var(R_{11})=E[\underbrace{Var(R_{ij}|U_{i0},U_{0j})}_{=E[R_{ij}^2|U_{i0},U_{0j}]}]+Var(\underbrace{E[R_{ij}|U_{i0},U_{0j}]}_{=0})=E[E[R_{ij}^2|U_{i0},U_{0j}]].
 \end{align*}
 In addition, the sum with $W_{ij}$ are asymptotically negligible to the rest.
 Under Assumption \ref{asm:MEL}, one can apply (\ref{pf:var-form}) and WLLN for triangular arrays \citep[Lemma 11.4.2]{lehmann2005testing} conditionally on $(U_{i0},U_{0j}:i=1,...,N,j=1,...M)$ to establish that the sum involving quadratic terms $R_{ij}$ and $W_{ij}$ is consistent for $\lim_{n\to \infty}2\sigma_{R}^2/NM(\sigma_{L}^2/n + \sigma_{R}^2/NM)$. Finally, we 
 note that the cross-product terms between the linear components and the quadratic components are asymptotically negligible. This verifies (\ref{pf:step3-1}) and thus concludes the proof.
}
%

\subsection{Proof of Theorem \ref{thm:mMEL}}

For brevity, we focus on the case where $X_{11}$ is scalar. Also we present the proof only for the nearly degenerate case (i.e., $\sigma_{L}^{2}=O(1/\underline{n})$ and $\sum_{i=1}^{N}\sum_{j=1}^{M}W_{ij}=o_{p}(\sum_{i=1}^{N}\sum_{j=1}^{M}R_{ij})$), as the non-degenerate case can be shown similarly. As in Step 1 in the proof of Theorem \ref{thm:MEL}, one can obtain the asymptotic expansion 
\[
\ell^{m}(\theta)=\frac{\left\{ n^{-1/2}\suml V_{l}^{m}(\theta)\right\} ^{2}}{n^{-1}\suml V_{l}^{m}(\theta)^{2}}+o_{p}(1).
\]
Define $\sigma^{2}=\sigma_{L}^{2}/n+\sigma_{R}^{2}/(NM)$. A similar argument to Step 3 in the proof of Theorem \ref{thm:MEL} together with the consistency of $\hat{\theta}$ imply $(n^{2}\sigma^{2})^{-1}\sum_{l=1}^{n}V_{l}^{m}(\hat{\theta})^{2}\stackrel{p}{\to}\lim_{n\to\infty}(\underline{n}\sigma_{L}^{2}+2\sigma_{R}^{2})/(\underline{n}\sigma_{L}^{2}+\sigma_{R}^{2})$. It now suffices to show 
\begin{equation}
\frac{1}{n^{2}\sigma^{2}}\sum_{l=1}^{N}\sum_{l_{1}=1}^{M}Q_{ll_{1}}^{2}=\frac{\sigma_{R}^{2}}{NM\sigma^{2}}+o_{p}(1),\label{pf:QQ}
\end{equation}
as the desired result is then implied by Step 3 in the proof of Theorem \ref{thm:MEL} as well as 
\begin{eqnarray*}
\frac{1}{n\sigma}\sum_{l=1}^{n}V_{l}^{m}(\theta) & = & \left\{ \frac{\sum_{l=1}^{n}V_{l}(\hat{\theta})^{2}}{\sum_{l=1}^{n}V_{l}(\hat{\theta})^{2}-\sum_{l=1}^{N}\sum_{l_{1}=1}^{M}Q_{ll_{1}}^{2}}\right\} ^{1/2}\frac{1}{n\sigma}\sum_{l=1}^{n}V_{l}(\theta)\\
 & = & \left\{ \frac{\underline{n}\sigma_{L}^{2}+2\sigma_{R}^{2}}{\underline{n}\sigma_{L}^{2}+\sigma_{R}^{2}}\right\} ^{1/2}\frac{1}{n\sigma}\sum_{l=1}^{n}V_{l}(\theta)+o_{p}(1).
\end{eqnarray*}
Notice that for each $l=1,\ldots,N$ and $l_{1}=1,\ldots,M$, eq. (\ref{eq:Hoeff}) yields 
\begin{eqnarray*}
nS(\theta) & = & \frac{n}{N}\sum_{i=1}^{N}L_{i0}+\frac{n}{M}\sum_{j=1}^{M}L_{0j}+\frac{n}{NM}\sum_{i=1}^{N}\sum_{j=1}^{M}\{W_{ij}+R_{ij}\},\\
(n-1)S_{l}(\theta) & = & \frac{n-1}{N-1}\sum_{i\neq l}^NL_{i0}+\frac{n-1}{M}\sum_{j=1}^{M}L_{0j}+\frac{n-1}{(N-1)M}\sum_{i\neq l}^N\sum_{j=1}^{M}\{W_{ij}+R_{ij}\},\\
(n-1)S_{N+l_{1}}(\theta) & = & \frac{n-1}{N}\sum_{i=1}^{N}L_{i0}+\frac{n-1}{M-1}\sum_{j\neq l_{1}}^M L_{0j}+\frac{n-1}{N(M-1)}\sum_{i=1}^{N}\sum_{j\neq l_{1}}^{M}\{W_{ij}+R_{ij}\},\\
(n-2)S_{l,l_{1}}(\theta) & = & \frac{n-2}{N-1}\sum_{i\neq l}^N L_{i0}+\frac{n-2}{M-1}\sum_{j\neq l_{1}}^ML_{0j}+\frac{n-2}{(N-1)(M-1)}\sum_{i\neq l}^N\sum_{j\neq l_{1}}^M\{W_{ij}+R_{ij}\}.
\end{eqnarray*}
By plugging in these expressions, a direct calculation yields
\begin{eqnarray}
Q_{ll_{1}} &=&\mathcal C(N,M)
\left\{-\frac{1}{N-1}\left(L_{l0}-\frac{1}{N}\sum_{i=1}^N L_{i0}\right)
-\frac{1}{M-1}\left(L_{0l_1}-\frac{1}{M}\sum_{j=1}^M L_{0j}\right)\right\} \nonumber \\
&& - \frac{1}{M^2}\sum_{j=1}^M(W_{lj}+R_{lj})
-\frac{1}{N^2}\sum_{i=1}^N (W_{il_1}+R_{il_1})
+\frac{n}{NM} (W_{ll_1}+R_{ll_1})+o_p(1). \label{pf:Qll1}
\end{eqnarray}
By inserting this expression to the left hand side of (\ref{pf:QQ}), applications of the weak law of large numbers under Assumption \ref{asm:MEL} (ii)-(iii) imply (\ref{pf:QQ}).

\subsection{Proof of Theorem \ref{thm:high}}
{  Here we provide a sketch of the proof; see the Online Supplementary Appendix for full details. }
By proceeding as in Chapter 2 of \cite{Hall1992}, we have
\[
\Pr\{\sqrt{T(\theta)}\le t\}=\Phi(t)-\left\{ (\kappa_{2,n}-1)t+\frac{\kappa_{4,n}}{12}(t^{3}-3t)\right\} \phi(t)+o((NM)^{-1/2}),
\]
where $\kappa_{q,n}$ is the $q$-th cumulant of $T(\theta)$. By an expansion, $T(\theta)$ can be written as 
\begin{equation}
T(\theta)=A\{1+B_{1}+B_{2}+B_{3}+B_{4}+B_{5}\}+o_{p}((NM)^{-1/2})\label{pf:T-exp}
\end{equation}
where 
\begin{eqnarray*}
A & = & \frac{1}{\sqrt{NM}}\sum_{a=1}^{N}\sum_{b=1}^{M}R_{ab},\quad B_{1}=-\frac{1}{2}\frac{1}{NM}\sum_{i=1}^{N}\sum_{k=1}^{M}\sum_{l\neq k}^{M}R_{ik}R_{il},\quad B_{2}=-\frac{1}{2}\frac{1}{NM}\sum_{i=1}^{N}\sum_{j\neq i}^{N}\sum_{k=1}^{M}R_{ik}R_{jk},\\
B_{3} & = & -\frac{1}{2}\left(\frac{1}{NM}\sum_{i=1}^{N}\sum_{k=1}^{M}R_{ik}^{2}-\sigma_{R}^{2}\right),\quad B_{4}=\frac{1}{2}\frac{N+M-1}{NM}\left(1+\frac{1}{NM}\sum_{i=1}^{N}\sum_{j\neq i}^{N}\sum_{k=1}^{M}\sum_{l\neq k}^{M}R_{ik}R_{jl}\right),\\
B_{5} & = & \frac{3}{8}\left(\frac{1}{NM}\sum_{i=1}^{N}\sum_{k=1}^{M}\sum_{l\neq k}^{M}R_{ik}R_{il}+\frac{1}{NM}\sum_{i=1}^{N}\sum_{j\neq i}^{N}\sum_{k=1}^{M}R_{ik}R_{jk}\right)^{2}.
\end{eqnarray*}
Then tedious but direct calculations yield
\begin{eqnarray*}
\kappa_{1,n} & = & -\frac{1}{2}(NM)^{-1/2}E[R_{ik}^{3}]+o((NM)^{-1/2}),\quad\kappa_{2,n}=1+\frac{3}{N}+\frac{3}{M}+o((NM)^{-1/2}),\\
\kappa_{3,n} & = & -2(NM)^{-1/2}E[R_{ij}^{3}]+o((NM)^{-1/2}),\quad\kappa_{4,n}=\frac{6}{N}+\frac{6}{M}+o((NM)^{-1/2}),
\end{eqnarray*}
and the conclusion for $T(\theta)$ follows.

We next consider the modified MEL statistic $\ell^{m}(\theta)$. 
By proceeding as in \cite{diccicio91} with $\sigma_{R}^{2}=1$, we obtain the signed root expansion
\begin{equation}
\sqrt{\ell^{m}(\theta)}=A\{1+B_{1}+B_{2}+B_{3}+B_{4}+B_{5}+C\}+o_{p}((NM)^{-1/2}),\label{pf:L-exp}
\end{equation}
where
\[
C=-\frac{1}{2}\left(\frac{3}{N}+\frac{3}{M}-\frac{5}{n}\right)+\frac{E[R_{ij}^{3}]}{3\sqrt{2}}\frac{1}{\sqrt{NM}}A-\left(\frac{\sqrt{2}-1}{2}\right)\frac{1}{n}A^{2}.
\]
Then the cumulants of $\sqrt{\ell^{m}(\theta)}$ are obtained as
\begin{eqnarray*}
\kappa_{1,n}^{m} & = & -\frac{1}{2}\frac{1}{\sqrt{NM}}E[R_{ij}^{3}]+o((NM)^{-1/2}),\quad\kappa_{2,n}^{m}=1+\frac{8-3\sqrt{2}}{n}+o((NM)^{-1/2}),\\
\kappa_{3,n}^{m} & = & -\frac{2}{\sqrt{NM}}E[R_{ij}^{3}]+o((NM)^{-1/2}),\quad\kappa_{4,n}^{m}=\frac{6}{N}+\frac{6}{M}-\frac{12(\sqrt{2}-1)}{n}+o((NM)^{-1/2}).
\end{eqnarray*}
By proceeding as in Chapter 2 of \cite{Hall1992}, we have
\[
\Pr\{\sqrt{\ell^{m}(\theta)}\le t\}=\Phi(t)-\left\{ (\kappa_{2,n}^{m}-1)t+\frac{\kappa_{4,n}^{m}}{12}(t^{3}-3t)\right\} \phi(t)+o((NM)^{-1/2}),
\]
and the conclusion for $\sqrt{\ell^{m}(\theta)}$ follows.

\subsection{Proof of Theorem \ref{thm:net}}

Here we focus on the case of $\theta=(\alpha,\beta^{\prime})^{\prime}$. The case where $\theta$ is a subvector of $(\alpha,\beta^{\prime})^{\prime}$ can be shown analogously by applying the argument in \cite{qin1994empirical}. Let $\hat{\theta}=(\hat{\alpha},\hat{\beta}^{\prime})^{\prime}$, $D_{ij}=(1,Z_{ij}^{\prime})^{\prime}$, $s_{ij}(\theta)=(Y_{ij}-\Lambda(D_{ij}^{\prime}\theta))D_{ij}$, and 
\[
H_{n}(\theta)=\frac{1}{NM}\sum_{i=1}^{N}\sum_{j=1}^{M}\nabla_{\theta}s_{ij}(\theta).
\]
\textbf{Step 1.} In this step, we shall derive the asymptotic distribution of $S(\theta)=\hat{\theta}-\theta$. Following Section 3 in \citet{Graham2020logit}, using the first-order condition for $\hat{\theta}$ and a mean value expansion yield that for some $\overline{\theta}$ that has each of its components lie between the corresponding components of $\hat{\theta}$ and $\theta$, it holds 
\[
S(\theta)=\hat{\theta}-\theta=\{nH_{n}(\bar{\theta})\}^{-1}\frac{n}{NM}\sum_{i=1}^{N}\sum_{j=1}^{M}s_{ij}(\theta).
\]
For the inverse factor in this expression, Appendix A of \citet{Graham2020logit} under Assumption \ref{asm:net} gives $\{nH_{n}(\bar{\theta})\}^{-1}\stackrel{p}{\to}H^{-1}$. For the linear component, we need further notation. Define $U_{i0}=(W_{i0},A_{i0})$, $U_{0j}=(W_{0j},A_{0j})$, and 
\begin{eqnarray*}
L_{i0} & = & E[s_{i1}(\theta_{0})|U_{i0}],\qquad L_{0j}=E[s_{1j}(\theta_{0})|U_{0j}],\\
W_{ij} & = & E[s_{ij}(\theta_{0})|U_{i0},U_{0j}]-E[s_{i1}(\theta_{0})|U_{i0}]-E[s_{1j}(\theta_{0})|U_{0j}],\\
R_{ij} & = & s_{ij}(\theta_{0})-E[s_{ij}(\theta_{0})|U_{i0},U_{0j}],\\
\sigma_{L}^{2} & = & n\{Var(L_{10})/N+Var(L_{01})/M\},\qquad\sigma_{R}^{2}=Var(R_{11})\\
\Sigma & = & \lim_{n\to\infty}n^{2}H^{-1}\{\sigma_{L}^{2}+\sigma_{R}^{2}/\underline{n}\}H^{-1},\qquad\Sigma_{m}=\lim_{n\to\infty}n^{2}H^{-1}\{\sigma_{L}^{2}+2\sigma_{R}^{2}/\underline{n}\}H^{-1}.
\end{eqnarray*}
Under Assumption \ref{asm:net}, by Theorem 1 in \citet{Graham2020logit} and the discussion there-before, we obtain 
\begin{eqnarray}
\sqrt{n}S(\theta) & = & \frac{n^{3/2}}{NM}H^{-1}\sum_{i=1}^{N}\sum_{j=1}^{M}s_{ij}(\theta)+o_{p}(1)\nonumber \\
 & = & n^{3/2}H^{-1}\left\{ \frac{1}{N}\sum_{i=1}^{N}L_{i0}+\frac{1}{M}\sum_{j=1}^{M}L_{0j}+\frac{1}{NM}\sum_{i=1}^{N}\sum_{j=1}^{M}R_{ij}\right\} +o_{p}(1)\nonumber\\
 &\overset{d}{\to}&N(0,\Sigma),\label{pf:net_Hoeff}
\end{eqnarray}
as well as that $n^{3/2}(NM)^{-1}H^{-1}\sum_{i=1}^{N}\sum_{j=1}^{M}W_{ij}=o_{p}(1)$.\\

\noindent \textbf{Step 2.} In this step, we shall prove
\begin{equation}
\frac{1}{\sqrt{n}}\sum_{l=1}^{n}V_{l}(\theta)=\sqrt{n}S(\theta)+o_{p}(1).\label{pf:net1}
\end{equation}
Since 
\[
\frac{1}{\sqrt{n}}\sum_{l=1}^{n}V_{l}(\theta)=\sqrt{n}S(\theta)-\frac{n-1}{\sqrt{n}}\sum_{l=1}^{n}\{S_{l}(\theta)-S(\theta)\},
\]
it is sufficient for (\ref{pf:net1}) to show $\sum_{l=1}^{n}S_{l}(\theta)=nS(\theta)+o_{p}(n^{-1/2})$. By a fourth-order Taylor expansion, it holds 
\begin{eqnarray}
S(\theta) & = & \hat{\theta}-\theta\nonumber \\
 & = & \{nH_{n}(\theta)\}^{-1}\frac{n}{NM}\sum_{i=1}^{N}\sum_{j=1}^{M}\Bigg\{ s_{ij}(\theta)+\sum_{k=1}^{d}\frac{\nabla_{\theta\theta_{k}}s_{ij}(\theta)}{2}(\hat{\theta}-\theta)(\hat{\theta}_{k}-\theta_{k})\nonumber \\
 &  & \qquad+\sum_{k=1}^{d}\sum_{\ell=1}^{d}\frac{\nabla_{\theta\theta_{k}\theta_{\ell}}s_{ij}(\theta)}{3!}(\hat{\theta}-\theta)(\hat{\theta}_{k}-\theta_{k})(\hat{\theta}_{\ell}-\theta_{\ell})\nonumber \\
 &  & \qquad+\sum_{k=1}^{d}\sum_{\ell=1}^{d}\sum_{m=1}^{d}\frac{\nabla_{\theta\theta_{k}\theta_{\ell}\theta_{m}}s_{ij}(\tilde{\theta})}{4!}(\hat{\theta}-\theta)(\hat{\theta}_{k}-\theta_{k})(\hat{\theta}_{\ell}-\theta_{\ell})(\hat{\theta}_{m}-\theta_{m})\Bigg\}\nonumber \\
 & = & \{nH_{n}(\theta)\}^{-1}\frac{n}{NM}\sum_{i=1}^{N}\sum_{j=1}^{M}\Bigg\{ s_{ij}(\theta)+\sum_{k=1}^{d}\frac{\nabla_{\theta\theta_{k}}s_{ij}(\theta)}{2}(\hat{\theta}-\theta)(\hat{\theta}_{k}-\theta_{k})\nonumber \\
 &  & \qquad+\sum_{k=1}^{d}\sum_{\ell=1}^{d}\frac{\nabla_{\theta\theta_{k}\theta_{\ell}}s_{ij}(\theta)}{3!}(\hat{\theta}-\theta)(\hat{\theta}_{k}-\theta_{k})(\hat{\theta}_{\ell}-\theta_{\ell})\Bigg\}+O_{p}(n^{-2})\nonumber \\
 & = & \frac{1}{NM}\sum_{i=1}^{N}\sum_{j=1}^{M}X_{ij}+O_{p}(n^{-2}),\label{eq:net_S}
\end{eqnarray}
for $\tilde{\theta}=(\tilde{\theta}_{k})_{k=1}^{d}$ with each $\tilde{\theta}_{k}$ lies between $\hat{\theta}_{k}$ and $\theta_{k}$, where 
\[
X_{ij}=n\{nH_{n}(\theta)\}^{-1}\left(s_{ij}(\theta)+\sum_{k=1}^{d}\frac{\nabla_{\theta\theta_{k}}s_{ij}(\theta)}{2}(\hat{\theta}-\theta)(\hat{\theta}_{k}-\theta_{k})+\sum_{k=1}^{d}\sum_{\ell=1}^{d}\frac{\nabla_{\theta\theta_{k}\theta_{\ell}}s_{ij}(\theta)}{3!}(\hat{\theta}-\theta)(\hat{\theta}_{k}-\theta_{k})(\hat{\theta}_{\ell}-\theta_{\ell})\right).
\]

We shall now establish similar asymptotic representations for the leave one column or row out counterparts. Denote 
\begin{eqnarray*}
H_{n-1}^{l}(\theta) & = & \begin{cases}
\frac{1}{(N-1)M}\sum_{i\ne l}\sum_{j=1}^{M}\nabla_{\theta}s_{ij}(\theta), & \text{if }l\le N,\\
\frac{1}{N(M-1)}\sum_{i=1}^{N}\sum_{j\ne l}\nabla_{\theta}s_{ij}(\theta), & \text{otherwise}.
\end{cases}
\end{eqnarray*}
For each $l\le N$ , we have 
\begin{eqnarray}
S_{l}(\theta) & = & \hat{\theta}^{(l)}-\theta\nonumber \\
 & = & \{nH_{n-1}^{l}(\theta)\}^{-1}\frac{n}{(N-1)M}\sum_{i\ne l}\sum_{j=1}^{M}\Bigg\{ s_{ij}(\theta)+\sum_{k=1}^{d}\frac{\nabla_{\theta\theta_{k}}s_{ij}(\theta)}{2}(\hat{\theta}^{(l)}-\theta)(\hat{\theta}_{k}^{(l)}-\theta_{k})\nonumber \\
 &  & \qquad+\sum_{k=1}^{d}\sum_{\ell=1}^{d}\frac{\nabla_{\theta\theta_{k}\theta_{\ell}}s_{ij}(\theta)}{3!}(\hat{\theta}^{(l)}-\theta)(\hat{\theta}_{k}^{(l)}-\theta_{k})(\hat{\theta}_{\ell}^{(l)}-\theta_{\ell})\nonumber \\
 &  & \qquad+\sum_{k=1}^{d}\sum_{\ell=1}^{d}\sum_{m=1}^{d}\frac{\nabla_{\theta\theta_{k}\theta_{\ell}\theta_{m}}s_{ij}(\tilde{\theta}^{(l)})}{4!}(\hat{\theta}^{(l)}-\theta)(\hat{\theta}_{k}^{(l)}-\theta_{k})(\hat{\theta}_{\ell}^{(l)}-\theta_{\ell})(\hat{\theta}_{m}^{(l)}-\theta_{m})\Bigg\}\nonumber \\
 & = & \{nH_{n}(\theta)\}^{-1}\frac{n}{(N-1)M}\sum_{i\ne l}\sum_{j=1}^{M}\Bigg\{ s_{ij}(\theta)+\sum_{k=1}^{d}\frac{\nabla_{\theta\theta_{k}}s_{ij}(\theta)}{2}(\hat{\theta}^{(l)}-\theta)(\hat{\theta}_{k}^{(l)}-\theta_{k})\nonumber \\
 &  & \qquad+\sum_{k=1}^{d}\sum_{\ell=1}^{d}\frac{\nabla_{\theta\theta_{k}\theta_{\ell}}s_{ij}(\theta)}{3!}(\hat{\theta}^{(l)}-\theta)(\hat{\theta}_{k}^{(l)}-\theta_{k})(\hat{\theta}_{\ell}^{(l)}-\theta_{\ell})\Bigg\}+O_{p}(n^{-2})\nonumber \\
 &  & \qquad-\left(\{nH_{n}(\theta)\}^{-1}-\{nH_{n-1}^{l}(\theta)\}^{-1}\right)\frac{n}{(N-1)M}\sum_{i\ne l}\sum_{j=1}^{M}s_{ij}(\theta)(1+o_{p}(1)),\label{pf:net_Sl}
\end{eqnarray}
where $\tilde{\theta}^{(l)}$ are vectors of mean values. Note that the last term of (\ref{pf:net_Sl}) has an order of $O_{p}(n^{-2})$ following the fact that $(B^{-1}-A^{-1})=B^{-1}(A-B)A^{-1}$ for $A$ and $B$ invertible, the fact that $nH_{n}(\theta)$ and $nH_{n-1}(\theta)$ have their eigenvalues bounded and bounded away from zero with probability approaching one as $H$ is of full rank, the asymptotic order $(n/NM)\sum_{i\ne l}\sum_{j=1}^{M}s_{ij}(\theta)=O_{p}(n^{-1/2})$ implied by the asymptotic normality in Step 1, and the claim that $n(H_{n}(\theta)-H_{n-1}^{l}(\theta))=O_{p}(n^{-3/2})$, which we will prove next. To show this claim, note that for each $l\le N$,
\begin{eqnarray}
 &  & n(H_{n}(\theta)-H_{n-1}^{l}(\theta))=-\frac{n}{NM}\sum_{i=1}^{N}\sum_{j=1}^{M}\nabla_{\theta}s_{ij}(\theta)+\frac{n}{(N-1)M}\sum_{i\ne l}\sum_{j=1}^{M}\nabla_{\theta}s_{ij}(\theta)\nonumber \\
 & = & -\frac{n}{NM}\sum_{i=1}^{N}\sum_{j=1}^{M}\nabla_{\theta}s_{ij}(\theta)+\frac{n}{(N-1)M}\sum_{i=1}^{N}\sum_{j=1}^{M}\nabla_{\theta}s_{ij}(\theta)-\frac{n}{(N-1)M}\sum_{j=1}^{M}\nabla_{\theta}s_{lj}(\theta)\nonumber \\
 & = & -\frac{1}{N-1}\left(\frac{1}{M}\sum_{j=1}^{M}\{n\nabla_{\theta}s_{lj}(\theta)-H\}-\frac{1}{NM}\sum_{i=1}^{N}\sum_{j=1}^{M}\{n\nabla_{\theta}s_{ij}(\theta)-H\}\right)\nonumber \\
 & = & O_{p}(n^{-1})O_{p}(n^{-1/2})=O_{p}(n^{-3/2}),\label{pf:net_bound_inverse_LOO}
\end{eqnarray}
where the last equality follows from applications of the maximal inequalities for separately exchangeable arrays \citep[Corollary 3]{ChiangKatoSasaki2020} and for i.i.d. random variables \citep[Theorem 2.14.1]{vdVW1996}, as well as the fact that $nE[\nabla_{\theta}s_{11}(\theta)]=H$. Hence (\ref{pf:net_Sl}) becomes 
\begin{eqnarray}
\hat{\theta}^{(l)}-\theta & = & \{nH_{n}(\theta)\}^{-1}\frac{n}{(N-1)M}\sum_{i\ne l}\sum_{j=1}^{M}\Bigg\{ s_{ij}(\theta)+\sum_{k=1}^{d}\frac{\nabla_{\theta\theta_{k}}s_{ij}(\theta)}{2}(\hat{\theta}^{(l)}-\theta)(\hat{\theta}_{k}^{(l)}-\theta_{k})\nonumber \\
 &  & \qquad+\sum_{k=1}^{d}\sum_{\ell=1}^{d}\frac{\nabla_{\theta\theta_{k}\theta_{\ell}}s_{ij}(\theta)}{3!}(\hat{\theta}^{(l)}-\theta)(\hat{\theta}_{k}^{(l)}-\theta_{k})(\hat{\theta}_{\ell}^{(l)}-\theta_{\ell})\Bigg\}+O_{p}(n^{-2})\nonumber \\
 & = & \{nH_{n}(\theta)\}^{-1}\frac{n}{(N-1)M}\sum_{i\ne l}\sum_{j=1}^{M}\Bigg\{ s_{ij}(\theta)+\sum_{k=1}^{d}\frac{\nabla_{\theta\theta_{k}}s_{ij}(\theta)}{2}(\hat{\theta}-\theta)(\hat{\theta}_{k}-\theta_{k})\nonumber \\
 &  & \qquad+\sum_{k=1}^{d}\sum_{\ell=1}^{d}\frac{\nabla_{\theta\theta_{k}\theta_{\ell}}s_{ij}(\theta)}{3!}(\hat{\theta}-\theta)(\hat{\theta}_{k}-\theta_{k})(\hat{\theta}_{\ell}-\theta_{\ell})\Bigg\}\nonumber\\
 &  &\qquad+O_{p}(n^{-2})+O_{p}(1)\cdot\left|\|\hat{\theta}^{(l)}-\theta\|^{2}-\|\hat{\theta}-\theta\|^{2}\right|\nonumber \\
 & = & \frac{1}{(N-1)M}\sum_{i\ne l}\sum_{j=1}^{M}X_{ij}+O_{p}(n^{-2})+O_{p}(1)\cdot\left|\|\hat{\theta}^{(l)}-\theta\|^{2}-\|\hat{\theta}-\theta\|^{2}\right|.\label{pf:net_Sl_recursive}
\end{eqnarray}
To obtain a concrete bound for the asymptotic order in (\ref{pf:net_Sl_recursive}), note that by the fact that $a^{2}-b^{2}=(a+b)(a-b)$, we have 
\begin{align}
\left|\|\hat{\theta}^{(l)}-\theta\|^{2}-\|\hat{\theta}-\theta\|^{2}\right|\le & \left|\|\hat{\theta}^{(l)}-\theta\|-\|\hat{\theta}-\theta\|\right|\cdot\left|\|\hat{\theta}^{(l)}-\theta\|+\|\hat{\theta}-\theta\|\right|\nonumber \\
\le & \|\hat{\theta}^{(l)}-\hat{\theta}\|\cdot\left|\|\hat{\theta}^{(l)}-\theta\|+\|\hat{\theta}-\theta\|\right|=\|\hat{\theta}^{(l)}-\hat{\theta}\|\cdot O_{p}(n^{-1/2}),\label{eq:net_Sl_order}
\end{align}
where the second inequality is implied by the reverse triangle inequality for norms $|\|a\|-\|b\||\le\|a-b\|$. We now proceed by the following recursive argument. Observe that using the coarse representations implied by the earlier Taylor expansions, it holds that 
\begin{eqnarray*}
\hat{\theta}-\theta & = & \{nH_{n}(\theta)\}^{-1}\frac{n}{NM}\sum_{i=1}^{N}\sum_{j=1}^{M}s_{ij}(\theta)+O_{p}(n^{-1}),\\
\hat{\theta}^{(l)}-\theta & = & \begin{cases}
\{nH_{n}(\theta)\}^{-1}\frac{n}{(N-1)M}\sum_{i\ne l}\sum_{j=1}^{M}s_{ij}(\theta)+O_{p}(n^{-1}) & \text{if }l\le N,\\
\{nH_{n}(\theta)\}^{-1}\frac{n}{N(M-1)}\sum_{i=1}^{N}\sum_{j\ne l}s_{ij}(\theta)+O_{p}(n^{-1}) & \text{otherwise}.
\end{cases}
\end{eqnarray*}
Following the same argument as (\ref{pf:net_bound_inverse_LOO}), we have $\|\hat{\theta}^{(l)}-\hat{\theta}\|=O_{p}(n^{-3/2})+O_{p}(n^{-1})=O_{p}(n^{-1})$. This implies that (\ref{eq:net_Sl_order}) is of order $O_{p}(n^{-3/2})$ and thus (\ref{pf:net_Sl_recursive}) gives 
\[
\hat{\theta}^{(l)}-\theta=\frac{1}{(N-1)M}\sum_{i\ne l}\sum_{j=1}^{M}X_{ij}+O_{p}(n^{-3/2}).
\]
Now, using this representation and (\ref{eq:net_S}), again following the same argument as (\ref{pf:net_bound_inverse_LOO}), we have $\|\hat{\theta}^{(l)}-\hat{\theta}\|=O_{p}(n^{-3/2})$ and thus (\ref{eq:net_Sl_order}) becomes $O_{p}(n^{-2})$. Plugging this into (\ref{pf:net_Sl_recursive}) once more, we have the representation
\begin{align*}
S_{l}(\theta)= & \hat{\theta}^{(l)}-\theta=\frac{1}{(N-1)M}\sum_{i\ne l}\sum_{j=1}^{M}X_{ij}+O_{p}(n^{-2}).
\end{align*}
Likewise, for $l>N$, a symmetric argument shows that 
\begin{align*}
S_{l}(\theta)= & \hat{\theta}^{(l)}-\theta=\frac{1}{N(M-1)}\sum_{i=1}^{N}\sum_{j\ne l}X_{ij}+O_{p}(n^{-2}).
\end{align*}
Utilizing these asymptotic representations yield that 
\begin{align*}
\sum_{l=1}^{n}S_{l}(\theta)  = & \frac{1}{(N-1)M}\sum_{l=1}^{N}\left(\sum_{i=1}^{N}\sum_{j=1}^{M}X_{ij}-\sum_{j=1}^{M}X_{lj}\right)+\frac{1}{N(M-1)}\sum_{l_{1}=1}^{M}\left(\sum_{i=1}^{N}\sum_{j=1}^{M}X_{ij}-\sum_{i=1}^{N}X_{il_{1}}\right)
+O_{p}(n\cdot n^{-2})\\
  = & \left\{ \frac{NM(N-1)}{(N-1)M}+\frac{NM(M-1)}{(M-1)N}\right\} S(\theta)+O_{p}(n^{-1})=nS(\theta)+o_{p}(n^{-1/2}),
\end{align*}
as required.\\

\noindent \textbf{Step 3.} In this step, we shall verify
\begin{align*}
\frac{1}{n}\sum_{l=1}^{n}V_{l}(\theta)V_{l}(\theta)^{\prime}\stackrel{p}{\to}\Sigma_{m},\qquad\frac{1}{n}\sum_{l=1}^{N}\sum_{l_{1}=1}^{M}Q_{ll_{1}}Q_{ll_{1}}^{\prime}\stackrel{p}{\to} & \lim_{n\to\infty}n^{2}H^{-1}(\sigma_{R}^{2}/\underline{n})H^{-1}.
\end{align*}
Using the Hoeffding type decomposition in (\ref{pf:net_Hoeff}), we have 
\begin{align*}
{\small S_{l}(\theta)-S_{l_{1}}(\theta)=\begin{cases}
\frac{nH^{-1}}{N-1}(L_{l_{1}0}-L_{l0})+\frac{nH^{-1}}{(N-1)M}\sum_{j=1}^{M}(R_{l_{1}j}-R_{lj})+o_{p}(n^{-1/2})\qquad\text{ if \ensuremath{l<l_{1}\le N}},\\
\frac{nH^{-1}}{M}L_{0(l_{1}-N)}-\frac{nH^{-1}}{N}L_{l0}+\frac{nH^{-1}}{N(M-1)}\sum_{i=1}^{N}R_{i(l_{1}-N)}-\frac{nH^{-1}}{(N-1)M}\sum_{j=1}^{M}R_{lj}+o_{p}(n^{-1/2})\:\text{if \ensuremath{l\le N<l_{1}}},\\
\frac{nH^{-1}}{M-1}(L_{0(l_{1}-N)}-L_{0(l-N)})+\frac{nH^{-1}}{N(M-1)}\sum_{i=1}^{N}(R_{i(l_{1}-N)}-R_{i(l-N)})+o_{p}(n^{-1/2})\quad\text{if \ensuremath{N<l<l_{1}}}.
\end{cases}}
\end{align*}
Using this and (\ref{pf:net1}), it holds 
\begin{eqnarray*}
 &  & \frac{1}{n}\sum_{l=1}^{n}V_{l}(\theta)V_{l}(\theta)^{\prime}=\frac{1}{n}\sum_{l=1}^{n}\{S(\theta)-(n-1)(S_{l}(\theta)-S(\theta))\}\{S(\theta)-(n-1)(S_{l}(\theta)-S(\theta))\}^{\prime}\\
 & = & \frac{(n-1)^{2}}{n}\sum_{l=1}^{n}\{S_{l}(\theta)-S(\theta)\}\{S_{l}(\theta)-S(\theta)\}^{\prime}+o_{p}(1)\\
 & = & \frac{(n-1)^{2}}{n^{2}}\sum_{l<l_{1}}\{S_{l}(\theta)-S_{l_{1}}(\theta)\}\{S_{l}(\theta)-S_{l_{1}}(\theta)\}^{\prime}+o_{p}(1)\\
 & = & \frac{2n^{2}}{N^{2}}\sum_{i<j\le N}\frac{H^{-1}(L_{i0}-L_{j0})(L_{i0}-L_{j0})^{\prime}H^{-1}}{2}+\frac{2n^{2}}{M^{2}}\sum_{i<j\le M}\frac{H^{-1}(L_{0i}-L_{0j})(L_{0i}-L_{0j})^{\prime}H^{-1}}{2}\\
 &  & \qquad+n^{2}\sum_{i=1}^{N}\sum_{j=1}^{M}H^{-1}\left(\frac{L_{0j}L_{0j}^{\prime}}{M^{2}}+\frac{L_{i0}L_{i0}^{\prime}}{N^{2}}\right)H^{-1}+\frac{n^{2}}{N^{2}M^{2}}\sum_{i<j\le N}\sum_{l=1}^{M}H^{-1}(R_{jl}-R_{il})(R_{jl}-R_{il})^{\prime}H^{-1}\\
 &  & \qquad+\frac{n^{2}}{N^{2}M^{2}}\sum_{i<j\le M}\sum_{k=1}^{N}H^{-1}(R_{kj}-R_{ki})(R_{kj}-R_{ki})^{\prime}H^{-1}\\
 &  & \qquad+\frac{n^{2}}{N^{2}M^{2}}\sum_{i=1}^{N}\sum_{j=1}^{M}H^{-1}\left(\sum_{k=1}^{N}R_{kj}-\sum_{l=1}^{M}R_{il}\right)\left(\sum_{k=1}^{N}R_{kj}-\sum_{l=1}^{M}R_{il}\right)^{\prime}H^{-1}+o_{p}(1)\\
 & = & n^{2}H^{-1}\left\{ \left(1+\frac{N}{M}\right)Var(L_{10})+\left(1+\frac{M}{N}\right)Var(L_{01})+\left(\frac{1}{M}+\frac{1}{N}+\frac{n}{NM}\right)Var(R_{11})\right\} H^{-1}+o_{p}(1)\\
 & \stackrel{p}{\to} & \Sigma_{m},
\end{eqnarray*}
where the last equality follows from the weak law of large numbers. On the other hand, using an analogous argument as in the proof of Theorem \ref{thm:mMEL}, we have 
\begin{eqnarray*}
 &  & \frac{1}{n}\sum_{l=1}^{N}\sum_{l_{1}=1}^{M}Q_{ll_{1}}Q_{ll_{1}}^{\prime}\\
 & = & \frac{1}{n}\sum_{l=1}^{N}\sum_{l_{1}=1}^{M}\left\{ \frac{n^2}{NM}H^{-1}R_{ll_{1}}+O_{p}(\underline{n}^{-3/2})\right\} \left\{ \frac{n^2}{NM}H^{-1}R_{ll_{1}}+O_{p}(\underline{n}^{-3/2})\right\} ^{\prime}+o_p(1)\\
 & \stackrel{p}{\to} & \lim_{n\to\infty}n^{2}H^{-1}(\sigma_{R}^{2}/\underline{n})H^{-1}.
\end{eqnarray*}

\noindent \textbf{Step 4.} Combining the results in Steps 1-3 with the same argument in the proof of Theorem \ref{thm:mMEL} and the fact that $\sum_{l=1}^{n}S_{l}(\hat{\theta})=\sum_{l=1}^{n}\{S_{l}(\theta)-S(\theta)\}$, we obtain 
\[
\frac{1}{\sqrt{n}}\sum_{l=1}^{n}V_{l}^{m}(\theta)\stackrel{d}{\to}N(0,\Sigma_{m}),\qquad\frac{1}{n}\sum_{l=1}^{n}V_{l}^{m}(\theta)V_{l}^{m}(\theta)^{\prime}\stackrel{p}{\to}\Sigma_{m}.
\]
The rest then follows straightforwardly from the linearization argument as in Step 1 in the proof of Theorem \ref{thm:MEL}.

\newpage{}

 \bibliographystyle{ecta}
\bibliography{MultiJEL}

\begin{thebibliography}{51}
\newcommand{\enquote}[1]{``#1''}
\expandafter\ifx\csname natexlab\endcsname\relax\def\natexlab#1{#1}\fi

\bibitem[\protect\citeauthoryear{Balan and Schiopu-Kratina}{Balan and
  Schiopu-Kratina}{2005}]{balan2005asymptotic}
\textsc{Balan, R.~M. and I.~Schiopu-Kratina} (2005): \enquote{Asymptotic
  results with generalized estimating equations for longitudinal data,}
  \emph{Annals of Statistics}, 33, 522--541.

\bibitem[\protect\citeauthoryear{Bertail}{Bertail}{2006}]{bertail2006empirical}
\textsc{Bertail, P.} (2006): \enquote{Empirical likelihood in some
  semiparametric models,} \emph{Bernoulli}, 12, 299--331.

\bibitem[\protect\citeauthoryear{Bhattacharyya and Bickel}{Bhattacharyya and
  Bickel}{2015}]{bhattacharyya2015subsampling}
\textsc{Bhattacharyya, S. and P.~J. Bickel} (2015): \enquote{Subsampling
  bootstrap of count features of networks,} \emph{Annals of Statistics}, 43,
  2384--2411.

\bibitem[\protect\citeauthoryear{Bickel, Chen, and Levina}{Bickel
  et~al.}{2011}]{BickelChenLevina2011}
\textsc{Bickel, P.~J., A.~Chen, and E.~Levina} (2011): \enquote{The method of
  moments and degree distributions for network models,} \emph{Annals of
  Statistics}, 39, 2280--2301.

\bibitem[\protect\citeauthoryear{Bravo, Escanciano, and Van~Keilegom}{Bravo
  et~al.}{2020}]{BravoEscancianoVan_Keilegom2020AoS}
\textsc{Bravo, F., J.~C. Escanciano, and I.~Van~Keilegom} (2020):
  \enquote{Two-step semiparametric empirical likelihood inference,}
  \emph{Annals of Statistics}, 48, 1--26.

\bibitem[\protect\citeauthoryear{Cameron, Gelbach, and Miller}{Cameron
  et~al.}{2011}]{CGM2011}
\textsc{Cameron, C.~A., J.~B. Gelbach, and D.~L. Miller} (2011):
  \enquote{Robust inference with multiway clustering,} \emph{Journal of
  Business \& Economic Statistics}, 29, 238--249.

\bibitem[\protect\citeauthoryear{Caron and Fox}{Caron and
  Fox}{2017}]{caron2017sparse}
\textsc{Caron, F. and E.~B. Fox} (2017): \enquote{Sparse graphs using
  exchangeable random measures,} \emph{Journal of the Royal Statistical
  Society: Series B (Statistical Methodology)}, 79, 1295.

\bibitem[\protect\citeauthoryear{Chen and Van~Keilegom}{Chen and
  Van~Keilegom}{2009}]{ChenVan_Keilegom2009Test}
\textsc{Chen, S.~X. and I.~Van~Keilegom} (2009): \enquote{A review on empirical
  likelihood methods for regression,} \emph{Test}, 18, 415--447.

\bibitem[\protect\citeauthoryear{Chiang, Kato, and Sasaki}{Chiang
  et~al.}{2021}]{ChiangKatoSasaki2020}
\textsc{Chiang, H.~D., K.~Kato, and Y.~Sasaki} (2021): \enquote{Inference for
  high-dimensional exchangeable arrays,} \emph{arXiv preprint
  arXiv:2009.05150}.

\bibitem[\protect\citeauthoryear{Chiang and Sasaki}{Chiang and
  Sasaki}{2023}]{chiang2023using}
\textsc{Chiang, H.~D. and Y.~Sasaki} (2023): \enquote{On Using The Two-Way
  Cluster-Robust Standard Errors,} \emph{arXiv preprint arXiv:2301.13775}.

\bibitem[\protect\citeauthoryear{Choi}{Choi}{2017}]{choi2017co}
\textsc{Choi, D.} (2017): \enquote{Co-clustering of nonsmooth graphons,}
  \emph{Annals of Statistics}, 45, 1488--1515.

\bibitem[\protect\citeauthoryear{Choi and Wolfe}{Choi and
  Wolfe}{2014}]{choi2014co}
\textsc{Choi, D. and P.~J. Wolfe} (2014): \enquote{Co-clustering separately
  exchangeable network data,} \emph{Annals of Statistics}, 42, 29--63.

\bibitem[\protect\citeauthoryear{Davezies, D'Haultfoeuille, and
  Guyonvarch}{Davezies et~al.}{2021}]{davezies2021}
\textsc{Davezies, L., X.~D'Haultfoeuille, and Y.~Guyonvarch} (2021):
  \enquote{Empirical process results for exchangeable arrays,} \emph{Annals of
  Statistics}, 49, 845--862.

\bibitem[\protect\citeauthoryear{Diaconis and Janson}{Diaconis and
  Janson}{2008}]{diaconis2008graph}
\textsc{Diaconis, P. and S.~Janson} (2008): \enquote{Graph limits and
  exchangeable random graphs,} \emph{Rendiconti di Matematica e delle sue
  Applicazioni}, 28, 33--61.

\bibitem[\protect\citeauthoryear{DiCiccio, Hall, and Romano}{DiCiccio
  et~al.}{1994}]{diccicio91}
\textsc{DiCiccio, T.~J., P.~Hall, and J.~Romano} (1994): \enquote{Empirical
  likelihood is Bartlett-correctable,} \emph{Annals of Statistics}, 19,
  1053--1061.

\bibitem[\protect\citeauthoryear{Efron and Stein}{Efron and
  Stein}{1981}]{EfronStein1981}
\textsc{Efron, B. and C.~Stein} (1981): \enquote{The jackknife estimate of
  variance,} \emph{Annals of Statistics}, 586--596.

\bibitem[\protect\citeauthoryear{Gao, Lu, and Zhou}{Gao
  et~al.}{2015}]{gao2015rate}
\textsc{Gao, C., Y.~Lu, and H.~H. Zhou} (2015): \enquote{Rate-optimal graphon
  estimation,} \emph{Annals of Statistics}, 43, 2624--2652.

\bibitem[\protect\citeauthoryear{Gong, Peng, and Qi}{Gong
  et~al.}{2010}]{GongPengQi2010JMA}
\textsc{Gong, Y., L.~Peng, and Y.~Qi} (2010): \enquote{Smoothed jackknife
  empirical likelihood method for ROC curve,} \emph{Journal of Multivariate
  Analysis}, 101, 1520--1531.

\bibitem[\protect\citeauthoryear{Graham}{Graham}{2020}]{Graham2020logit}
\textsc{Graham, B.~S.} (2020): \enquote{Sparse network asymptotics for logistic
  regression,} \emph{working paper}.

\bibitem[\protect\citeauthoryear{Graham, Niu, and Powell}{Graham
  et~al.}{2019}]{Graham2019kernel}
\textsc{Graham, B.~S., F.~Niu, and J.~L. Powell} (2019): \enquote{Kernel
  density estimation for undirected dyadic data,} \emph{arXiv preprint
  arXiv:1907.13630}.

\bibitem[\protect\citeauthoryear{Hall}{Hall}{1992}]{Hall1992}
\textsc{Hall, P.} (1992): \emph{The bootstrap and Edgeworth expansion},
  Springer.

\bibitem[\protect\citeauthoryear{Helmers}{Helmers}{1991}]{helmers1991edgeworth}
\textsc{Helmers, R.} (1991): \enquote{On the Edgeworth expansion and the
  bootstrap approximation for a Studentized U-statistic,} \emph{Annals of
  Statistics}, 470--484.

\bibitem[\protect\citeauthoryear{Hinkley}{Hinkley}{1978}]{hinkley1978improving}
\textsc{Hinkley, D.~V.} (1978): \enquote{Improving the jackknife with special
  reference to correlation estimation,} \emph{Biometrika}, 65, 13--21.

\bibitem[\protect\citeauthoryear{Hjort, McKeague, and Van~Keilegom}{Hjort
  et~al.}{2009}]{HjortMcKeagueVan_Keilegom2009AoS}
\textsc{Hjort, N.~L., I.~W. McKeague, and I.~Van~Keilegom} (2009):
  \enquote{Extending the scope of empirical likelihood,} \emph{Annals of
  Statistics}, 37, 1079--1111.

\bibitem[\protect\citeauthoryear{Jing, Yuan, and Zhou}{Jing
  et~al.}{2009}]{JingYuanZhang2009}
\textsc{Jing, B.-Y., J.~Yuan, and W.~Zhou} (2009): \enquote{Jackknife empirical
  likelihood,} \emph{Journal of the American Statistical Association}, 104,
  1224--1232.

\bibitem[\protect\citeauthoryear{Kallenberg}{Kallenberg}{2006}]{Kallenberg2006}
\textsc{Kallenberg, O.} (2006): \emph{Probabilistic Symmetries and Invariance
  Principles}, Springer Science \& Business Media.

\bibitem[\protect\citeauthoryear{Lauritzen, Rinaldo, and Sadeghi}{Lauritzen
  et~al.}{2018}]{lauritzen2018random}
\textsc{Lauritzen, S., A.~Rinaldo, and K.~Sadeghi} (2018): \enquote{Random
  networks, graphical models and exchangeability,} \emph{Journal of the Royal
  Statistical Society: Series B (Statistical Methodology)}, 80, 481--508.

\bibitem[\protect\citeauthoryear{Lehmann and Romano}{Lehmann and
  Romano}{2005}]{lehmann2005testing}
\textsc{Lehmann, E. and J.~P. Romano} (2005): \enquote{Testing Statistical
  Hypotheses,} \emph{Springer Texts in Statistics}.

\bibitem[\protect\citeauthoryear{Liang and Zeger}{Liang and
  Zeger}{1986}]{liang1986longitudinal}
\textsc{Liang, K.-Y. and S.~L. Zeger} (1986): \enquote{Longitudinal data
  analysis using generalized linear models,} \emph{Biometrika}, 73, 13--22.

\bibitem[\protect\citeauthoryear{MacKinnon, Nielsen, and Webb}{MacKinnon
  et~al.}{2021}]{mackinnon2021wild}
\textsc{MacKinnon, J.~G., M.~{\O}. Nielsen, and M.~D. Webb} (2021):
  \enquote{Wild bootstrap and asymptotic inference with multiway clustering,}
  \emph{Journal of Business \& Economic Statistics}, 39, 505--519.

\bibitem[\protect\citeauthoryear{Matsushita and Otsu}{Matsushita and
  Otsu}{2021}]{MatsushitaOtsu2020}
\textsc{Matsushita, Y. and T.~Otsu} (2021): \enquote{Jackknife empirical
  likelihood: small bandwidth, sparse network and high-dimension asymptotic,}
  \emph{Biometrika}, forthcoming.

\bibitem[\protect\citeauthoryear{McCullagh}{McCullagh}{2000}]{Mccullagh2000}
\textsc{McCullagh, P.} (2000): \enquote{Resampling and exchangeable arrays,}
  \emph{Bernoulli}, 6, 285--301.

\bibitem[\protect\citeauthoryear{Menzel}{Menzel}{2021}]{menzel2021bootstrap}
\textsc{Menzel, K.} (2021): \enquote{Bootstrap with cluster-dependence in two
  or more dimensions,} \emph{Econometrica}, 89, 2143--2188.

\bibitem[\protect\citeauthoryear{Miglioretti and Heagerty}{Miglioretti and
  Heagerty}{2007}]{miglioretti2007marginal}
\textsc{Miglioretti, D.~L. and P.~J. Heagerty} (2007): \enquote{Marginal
  modeling of nonnested multilevel data using standard software,}
  \emph{American Journal of Epidemiology}, 165, 453--463.

\bibitem[\protect\citeauthoryear{Orbanz and Roy}{Orbanz and
  Roy}{2014}]{orbanz2014bayesian}
\textsc{Orbanz, P. and D.~M. Roy} (2014): \enquote{Bayesian models of graphs,
  arrays and other exchangeable random structures,} \emph{IEEE Transactions on
  Pattern Analysis and Machine Intelligence}, 37, 437--461.

\bibitem[\protect\citeauthoryear{Owen}{Owen}{1988}]{Owen1988}
\textsc{Owen, A.~B.} (1988): \enquote{Empirical likelihood ratio confidence
  intervals for a single functional,} \emph{Biometrika}, 75, 237--249.

\bibitem[\protect\citeauthoryear{Owen}{Owen}{1990}]{Owen1990}
---\hspace{-.1pt}---\hspace{-.1pt}--- (1990): \enquote{Empirical likelihood
  ratio confidence regions,} \emph{Annals of Statistics}, 90--120.

\bibitem[\protect\citeauthoryear{Owen}{Owen}{2001}]{Owen2001Book}
---\hspace{-.1pt}---\hspace{-.1pt}--- (2001): \emph{Empirical Likelihood}, CRC
  press.

\bibitem[\protect\citeauthoryear{Owen}{Owen}{2007}]{Owen2007}
---\hspace{-.1pt}---\hspace{-.1pt}--- (2007): \enquote{The pigeonhole
  bootstrap,} \emph{Annals of Applied Statistics}, 1, 386--411.

\bibitem[\protect\citeauthoryear{Putter and van Zwet}{Putter and van
  Zwet}{1998}]{putter1998empirical}
\textsc{Putter, H. and W.~R. van Zwet} (1998): \enquote{Empirical Edgeworth
  expansions for symmetric statistics,} \emph{Annals of Statistics}, 26,
  1540--1569.

\bibitem[\protect\citeauthoryear{Qin and Lawless}{Qin and
  Lawless}{1994}]{qin1994empirical}
\textsc{Qin, J. and J.~Lawless} (1994): \enquote{Empirical likelihood and
  general estimating equations,} \emph{Annals of Statistics}, 22, 300--325.

\bibitem[\protect\citeauthoryear{Searle, Casella, and McCulloch}{Searle
  et~al.}{2009}]{searle2009variance}
\textsc{Searle, S.~R., G.~Casella, and C.~E. McCulloch} (2009): \emph{Variance
  Components}, John Wiley \& Sons.

\bibitem[\protect\citeauthoryear{Thompson}{Thompson}{2011}]{thompson2011simple}
\textsc{Thompson, S.~B.} (2011): \enquote{Simple formulas for standard errors
  that cluster by both firm and time,} \emph{Journal of Financial Economics},
  99, 1--10.

\bibitem[\protect\citeauthoryear{van~der Vaart and Wellner}{van~der Vaart and
  Wellner}{1996}]{vdVW1996}
\textsc{van~der Vaart, A.~W. and J.~A. Wellner} (1996): \emph{Weak Convergence
  and Empirical Processes}, Springer.

\bibitem[\protect\citeauthoryear{Veitch and Roy}{Veitch and
  Roy}{2019}]{veitch2019sampling}
\textsc{Veitch, V. and D.~M. Roy} (2019): \enquote{Sampling and estimation for
  (sparse) exchangeable graphs,} \emph{Annals of Statistics}, 47, 3274--3299.

\bibitem[\protect\citeauthoryear{White}{White}{2001}]{white2001asymptotic}
\textsc{White, H.} (2001): \emph{Asymptotic Theory for Econometricians},
  Academic press.

\bibitem[\protect\citeauthoryear{Xie and Yang}{Xie and
  Yang}{2003}]{xie2003asymptotics}
\textsc{Xie, M. and Y.~Yang} (2003): \enquote{Asymptotics for generalized
  estimating equations with large cluster sizes,} \emph{Annals of Statistics},
  31, 310--347.

\bibitem[\protect\citeauthoryear{Zhang, Levina, and Zhu}{Zhang
  et~al.}{2017}]{zhang2017estimating}
\textsc{Zhang, Y., E.~Levina, and J.~Zhu} (2017): \enquote{Estimating network
  edge probabilities by neighbourhood smoothing,} \emph{Biometrika}, 104,
  771--783.

\bibitem[\protect\citeauthoryear{Zhang and Zhao}{Zhang and
  Zhao}{2013}]{zhang2013empirical}
\textsc{Zhang, Z. and Y.~Zhao} (2013): \enquote{Empirical likelihood for linear
  transformation models with interval-censored failure time data,}
  \emph{Journal of Multivariate Analysis}, 116, 398--409.

\bibitem[\protect\citeauthoryear{Zhong and Chen}{Zhong and
  Chen}{2014}]{zhong2014jackknife}
\textsc{Zhong, P.-S. and S.~Chen} (2014): \enquote{Jackknife empirical
  likelihood inference with regression imputation and survey data,}
  \emph{Journal of Multivariate Analysis}, 129, 193--205.

\bibitem[\protect\citeauthoryear{Zhu and Xue}{Zhu and
  Xue}{2006}]{zhu2006empirical}
\textsc{Zhu, L. and L.~Xue} (2006): \enquote{Empirical likelihood confidence
  regions in a partially linear single-index model,} \emph{Journal of the Royal
  Statistical Society: Series B (Statistical Methodology)}, 68, 549--570.

\end{thebibliography}

\end{document}